\documentclass[journal]{IEEEtran}
\usepackage{amsmath,amsfonts}
\usepackage{algorithm}
\usepackage{algorithmicx}
 \usepackage{algpseudocode}
 \usepackage{array}
\usepackage[caption=false,font=normalsize,labelfont=sf,textfont=sf]{subfig}
\usepackage{textcomp}
\usepackage{stfloats}
\usepackage{url}
\usepackage{verbatim}
\usepackage{graphicx}
\usepackage{cite}
\usepackage{hyperref}
\usepackage{xcolor}
\usepackage{soul}
\graphicspath{ {./} }
\usepackage{xcolor}
\def\BibTeX{{\rm B\kern-.05em{\sc i\kern-.025em b}\kern-.08em
		T\kern-.1667em\lower.7ex\hbox{E}\kern-.125emX}}

\begin{document}
	
	\title{Adaptive Attack Mitigation for IoV Flood Attacks\thanks{Manuscript received XX, 2024; revised  YY, 2024.}\thanks{Copyright (c) 20xx IEEE. Personal use of this material is permitted. However, permission to use this material for any other purposes must be obtained from the IEEE by sending a request to pubs-permissions@ieee.org}}

	\author{Erol Gelenbe\thanks{Prof. Erol Gelenbe is with the Institute of Theoretical \& Applied Informatics, Polish Academy of Science, 44-100 Gliwice, PL, also an Honorary Researcher at
			CNRS I3S, Universit\'{e} C\^{o}te d'Azur, 06100 Nice, FR, and a Visiting Professor in the Department of Engineering, King's College London, Strand WC2R 2LS, UK.
			ORCID: 0000-0001-9688-2201, email: gelenbe.erol@gmail.com},~\IEEEmembership{Fellow,~IEEE,} and Mohammed Nasereddin\thanks{Mohammed Nasereddin is with the Institute of Theoretical \& Applied Informatics, Polish Academy of Science, 44-100 Gliwice, PL.
			ORCID: 0000-0002-3740-9518, email: mnasereddin@iitis.pl}}

	\markboth{Journal of \LaTeX\ Class Files,~Vol.~XX, No.~Y, ABCD~20ZZ}%
{Shell \MakeLowercase{\textit{et al.}}: A Sample Article Using IEEEtran.cls for IEEE Journals}
	

	\maketitle

	\begin{abstract}
Gateway Servers for the Internet of Vehicles (IoV) must meet stringent Security and Quality of Service (QoS) requirements, including cyberattack protection, low delays and minimal packet loss, to offer secure real-time data exchange for human and vehicle safety and efficient road traffic management. Therefore, it is vital to protect these systems from cyberattacks with adequate Attack Detection (AD) and Mitigation mechanisms. Such attacks often include packet Floods that impair the QoS of the networks and Gateways and even impede the Gateways' capability to carry out AD. Thus, this paper first evaluates these effects using system measurements during Flood attacks. It then demonstrates how a Smart Quasi-Deterministic Policy Forwarder (SQF) at the entrance of the Gateway can regulate the incoming traffic to ensure that the Gateway supports the AD to operate promptly during an attack. Since Flood attacks create substantial packet backlogs, we propose a novel Adaptive Attack Mitigation (AAM) system that is activated after an attack is detected to dynamically sample the incoming packet stream, determine whether the attack is continuing, and also drop batches of packets at the input to reduce the effects of the attack. The AAM is designed to minimize a cost function that includes the sampling overhead and the cost of lost benign packets. We show experimentally that the Optimum AAM approach is effective in mitigating attacks and present theoretical and experimental results that validate the proposed approach.
	\end{abstract}
	
	\begin{IEEEkeywords}
	Cyberattack Detection and Mitigation, Internet of Vehicles, Flood Attacks,  Quasi-Deterministic Transmission Policy, Adaptive Attack Mitigation
\end{IEEEkeywords}



	
	
	

	\section{Introduction}
	
Smart vehicles rely on many onboard and road-side IoT devices, and today's  $30$ Billion or more IoT devices \cite{Cisco2020} are already subjected to numerous cyberattacks
\cite{liu2017novel,tello2018performance} which disable systems with huge packet floods \cite{Cyber23}. For instance, an attack in $2017$  took down $180,000$ servers with an overall  $2.54$ Terabits per second of traffic \cite{Cloudflare1}. 
Thus,  it is imperative to develop the understanding and technology that can keep both the IoV and the IoT as safe as possible. 


Unlike the general IoT which mainly involves stationary, low power \cite{Sensor-Energy} or occasionally mobile devices, the IoV operates mostly with numerous continuously moving nodes, i.e. vehicles, that generate, process, and transmit data in real-time, together with road-side monitoring and relaying systems whose complexity may cause the random misdirection and loss of data \cite{gelenbe2007diffusion}. The safety-critical nature of the  IoV, which directly impacts road safety and traffic efficiency, demands stringent requirements on data timeliness, reliability, and very low latency \cite{Pokhrel,Savazzi}, making it particularly vulnerable to Flood attacks, which overload the network’s gateway servers that act as control hubs which manage the flow of data between vehicles and the broader network infrastructure, and severely compromise QoS.

Smart vehicles may number two billion by 2035 \cite{Voelcker_vehicles_stats}, so that supply chains and manufacturing networks will also rely in the future on smart transport \cite{Supply1}. Since smart vehicles depend on timely and accurate data collection and forwarding,  cyberattacks against smart vehicles can cause network delays, data loss and inaccuracy, resulting in road traffic gridlock, driver and passenger frustration, delayed delivery of goods and services, and wrong vehicle traffic control decisions \cite{Xu} that can lead to road congestion and impair safety  \cite{Frotscher}.

Therefore, the present paper proposes an Attack Mitigation approach that operates together with an Attack Detector (AD), that acts at the input of an IoV or IoT Gateway to detect attacks rapidly in real time to optimally mitigate their effects. The AD  is based on previous work on highly accurate Machine Learning (ML) attack detection methods \cite{brun2018deep,CDIS,nakip2024online} which determine whether the incoming traffic deviates from the expected normal behaviour. The novel Adaptive Attack Mitigation (AAM) scheme that we propose, rapidly drops incoming attack packets when the AD produces an alarm and actively samples the incoming packet stream in a manner that minimizes the Gateway overhead and the cost of mistakingly dropping benign packets.

\subsection{Related Work on the Security of the IoV and IoT} \label{Related}
	
The literature on cybersecurity for the IoV and the IoT, encompasses a range of studies addressing attack and defense methodologies, proposed solutions, and related challenges. Several security architectures, lightweight protocols, and frameworks for autonomous vehicle communication are discussed in \cite{kim2021cybersecurity}, emphasizing the integration of advanced security solutions. In \cite{banafshehvaragh2023intrusion}, attack and defense studies are categorized to shed  light on methods and to identify security challenges for future exploration.  In the realm of autonomous driving, the integration of artificial intelligence (AI) into networks and network control algorithms,  raises additional cybersecurity concerns that must be considered \cite{girdhar2023cybersecurity,man2021ai}.

Cyberattack detection algorithms may be trained offline with calibrated real or synthetic datasets, or online with real data. Most studies then evaluate the accuracy of the resulting trained ADs with real or synthetic datasets, using the following performance scores that are obtained from the total number of true positives (TP), true negatives (TN), false positives (FP) or false alarms, and false negatives (FN), obtained during testing:
\begin{eqnarray}
&&Accuracy =\frac{TP+TN}{TP+TN+FP+FN},\nonumber\\
&&True~Positive~Rate~(TPR)=\frac{TP}{TP+FN},\nonumber\\
&&True~Negative~Rate~(TNR)=\frac{TN}{TN+FP}.\nonumber
\end{eqnarray}
In some cases, other metrics such as the $F1$ and $F2$ scores are also used.

The research on ADs, their design
\cite{Douligeris} and accuracy
\cite{Oke,AHMETOGLU22}, often use ML methods that are  evaluated off-line  \cite{Banerjee,Tuan,Al-Issa,Guven} with various datasets, rather than with on-line traffic traces. Such evaluations typically neglect the overload created by attacks on the victim devices or systems, which  can paralyze the targeted system with floods of packets, in addition to compromising them with malware.

\begin{table*}[ht]
	\caption{Performance Comparison of Attack Detectors for IoV Networks}
	\centering
	\begin{tabular}{|p{3.25cm}|p{4.5cm}|p{3cm}|p{4.75cm}|}
		\hline
		\textbf{CyberAttack Detection for the IoV} & \textbf{Detection Method} & \textbf{Evaluation 
			Strategy} & \textbf{Evaluation Metrics: \newline Accuracy 
			$\mid$$\mid$ TPR $\mid$$\mid$ TNR} \\
		\hline
	``Secure attack detection framework for hierarchical 6G-enabled 
		internet of vehicles''  \cite{sedjelmaci2023secure}
		& Hybrid (Federated Learning (FL) + Stackelberg Game)
		& Simulation using the UNSW-NB15 dataset
		& \begin{tabular}[c]{@{}l@{}}
			$96\%$ $\mid$$\mid$
			$95\%$ $\mid$$\mid$
			$97\%$
		\end{tabular} \\
		\hline
		``A Transfer Learning and Optimized CNN Based Intrusion Detection 
		System for Internet of Vehicles'' \cite{yang2022transfer}
		& Deep Learning (Transfer Learning + Optimized CNN)
		& Car-Hacking and CICIDS2017 datasets
		& \begin{tabular}[c]{@{}l@{}}
			$99.93\%$ $\mid$$\mid$
			$99.90\%$ $\mid$$\mid$
			$99.82\%$
		\end{tabular} \\
		\hline
		``An Intrusion Detection System for Connected Vehicles in Smart Cities'' \cite{aloqaily2019intrusion}
		& Hybrid (Deep Belief Network (DBN)+ Decision Tree (DT))
		& Simulation using ns-3 and NSL-KDD datasets with validation through 
		MATLAB simulations
		& \begin{tabular}[c]{@{}l@{}}
			$99.43\%$ $\mid$$\mid$
			$99.04\%$ $\mid$$\mid$
			$98.47\%$
		\end{tabular} \\
		\hline
		``Cybersecurity for Autonomous Vehicles Against Malware Attacks in 
		Smart Cities'' \cite{aurangzeb2023cybersecurity}
		& Hybrid (Static + Dynamic Analysis with ML)
		& Evaluated using a custom dataset of $1000$ malware downloaded from 
		\textit{Virusshare.com} and $1000$ non-malware applications
		& \begin{tabular}[c]{@{}l@{}}
			$96.3\%$ $\mid$$\mid$
			$95.8\%$ $\mid$$\mid$
			$97.2\%$
		\end{tabular} \\
		\hline
		``Cybersecurity in Automotive: An Intrusion Detection System in 
		Connected Vehicles'' 	\cite{pascale2021cybersecurity}
		& Hybrid (Bayesian Networks + Spatial-Temporal Analysis)
		& Evaluated using simulated datasets generated through the CARLA 
		simulator for various attack scenarios
		& \begin{tabular}[c]{@{}l@{}}
			$98.2\%$ $\mid$$\mid$
			$97.8\%$ $\mid$$\mid$
			$98.5\%$
		\end{tabular} \\
		\hline
		``MTH-IDS: A multitiered hybrid intrusion detection system for internet 
		of vehicles''  \cite{yang2021mth}
		& Hybrid (Tree-based Ensemble Learning + Clustering)
		& Cross-validation on CAN-intrusion and CICIDS2017 datasets
		& \begin{tabular}[c]{@{}l@{}}
			$99.99\%$ (CAN) $\mid$$\mid$ $99.88\%$ (CICIDS2017) \\
			$99.81\%$ $\mid$$\mid$
			$99.89\%$			
		\end{tabular} \\
		\hline
		``Detection and identification of malicious cyber-attacks in connected 
		and automated vehicles' real-time sensors''  \cite{eziama2020detection} 
		& Hybrid (Discrete Wavelet Transform + Bayesian DL)
		& Evaluated on simulated data generated from the Safety Pilot Model 
		Deployment dataset
		& \begin{tabular}[c]{@{}l@{}}
			$98.3\%$ $\mid$$\mid$
			$97.9\%$ $\mid$$\mid$
			$98.1\%$
		\end{tabular} \\
		\hline
		\textbf{Our System in this Paper}
		& \textbf{Hybrid (Random Neural Network + Adaptive Attack 
			Mitigation)}
		& \textbf{Evaluated on a custom testbed with flood attacks 
			implemented using the MHDDoS public repository}
		& \begin{tabular}[c]{@{}l@{}}
			\textbf{99.71\% $\mid\mid$
				99.73\%$ \mid\mid$
				98.48\%}
		\end{tabular} \\
		\hline
	\end{tabular}
	\label{tab:attack_Table}
\end{table*}

Some cyberattack test-beds \cite{mirkovic2009test}  have been
constructed for experimentation \cite{Kaouk2018,annor2018development,waraga2020design}
in critical applications such as wind farms \cite{singh2020testbed}, and Supervisory Control and Data Acquisition (SCADA) systems \cite{ghaleb2016scada,tesfahun2016scada}.
Experiments with Flood attacks were presented in \cite{park2018test}, and real-time  DNS attack data was collected in \cite{arthi2021design}. Software Defined Networks used for smart network routing, that adapt the packet flows to the needs of smart vehicles, can also be subjected to Denial of Service (DoS) attacks \cite{wright2019testbed}. Attack emulation, without considering the effect of the overload that attacks produce, was investigated for autonomous vehicles in \cite{sontakke2022impact}.

ML and Deep Learning (DL) take center stage in many AD studies. For instance, a Unified Modeling Language (UML) based framework \cite{he2020machine}  introduces Decision Tree and Naive Bayes algorithms to classify vulnerabilities. In a bid to fortify vehicle network security, an innovative AI-based solution \cite{aldhyani2022attacks} employs Convolutional Neural Networks (CNN) and Hybrid CNN Long and Short-Term Memory (CNN-LSTM) models for the detection and classification of message attacks. Additionally, the abnormal behavior within vehicle networks has been studied with a classification that uses a generative adversarial network (GAN) in \cite{kavousi2020evolutionary}.
	
	Security concerns for the IoV have received further attention in \cite{ahmed2021deep} with DL to reduce false positives and enhance the resilience of the transportation ecosystem. In \cite{sharma2020machine,alladi2021securing},  data-centric misbehavior detection based on supervised ML and transfer learning has also been proposed. A holistic approach in \cite{sedjelmaci2023secure} presents a detection framework tailored for 6G-enabled IoV, with edge node processing, federated learning, and robust security measures.
	
	The IoT and IoV can also use different devices from different vendors that may not wish to share their network datasets. For instance, the healthcare IoT may use body sensors, hospital security devices, and patient monitoring, while the IoV may be used for distinct vehicles that convey patients. Thus recent research has also investigated cyberattack detection methods that use transfer learning between distinct IoT devices and distinct vehicles that do not wish to share data, but which need to take advantage of the cybersecurity experience of their peers \cite{DISFIDA}.	Transfer and ensemble learning using CNNs have also been combined to achieve impressive attack detection rates that exceed $99.25\%$ \cite{yang2022transfer}. In \cite{alladi2021artificial}, where DL is deployed on Multi-access Edge Computing (MEC) servers, demonstrating great efficacy in classifying potential cyberattacks and enhancing real-time security for IoV networks. Expanding the focus to connected vehicles in smart cities, an AD system utilizing spatial and temporal analysis and Bayesian networks is proposed in \cite{aloqaily2019intrusion}, while a hybrid approach combining static and dynamic analyses is discussed in \cite{aurangzeb2023cybersecurity}. In \cite{pascale2021cybersecurity,yang2021mth},  embedded AD systems for the automotive sector are presented. Another example combining Bayesian DL and Discrete Wavelet Transform for proactive detection can also be found in \cite{eziama2020detection}.
	
	Since many types of attacks may target the Authentication, Availability, Secrecy, Routing, and Data Authenticity in automobile networks \cite{sun2017attacks,yang2019tree}, the dynamic nature of the IoV and the presence of varied cyberattacks have also been addressed with transfer learning using cloud-assisted updates \cite{li2021transfer}.
	
In addition to the previous references, seven state-of-the-art attack detection methods for IoV attacks are discussed in detail and compared with the method that we use in this paper in Table \ref{tab:attack_Table}.	Furthermore, five recent mitigation techniques for IoV attacks are also discussed and compared with the method that we have developed in this paper in Table \ref{tab:mitigation_Table}.
	
	\begin{table*}[ht]
		\caption{Comparison of Mitigation Approaches in IoV Networks}
		\centering
		\resizebox{\textwidth}{!}{%
			\begin{tabular}{|p{2.5cm}|p{2.5cm}|p{4.5cm}|p{3.5cm}|p{4cm}|p{4cm}|}
				\hline
				\textbf{Reference} & \textbf{Mitigation Approach} & \textbf{Brief 
					Description} & \textbf{Evaluation Strategy} & \textbf{Evaluation 
					Results} & \textbf{Limitations} \\
				\hline
				``A Comprehensive Detection and Mitigation Mechanism to Protect SD-IoV 
				Against DDoS Attacks ''\cite{TM1} & Hybrid (Entropy + Flow Rate Analysis) & Utilizes 
				entropy metrics to detect anomalies by analyzing payload lengths and 
				mitigates attacks using adaptive thresholds & Simulated using SUMO, 
				Mininet-WiFi, and Scapy on a SD-IoV environment with different scenarios 
				& Reduced network load and bandwidth congestion by blocking IPs that 
				send a number of packets exceeding the identified threshold for a 
				specified period, before re-evaluating them & May not scale for 
				high-volume traffic, as entropy and flow rate analysis over sliding 
				windows can be computationally intensive, causing real-time delays 
				\textbf{$\mid$\&$\mid$} Adaptive thresholds require careful tuning to 
				handle varying network conditions to avoid false positives \\
				\hline
				``A Novel DDoS Mitigation Strategy in 5G-Based Vehicular 
				Networks Using Chebyshev Polynomials'' \cite{TM2} & Chebyshev Polynomials-based & 
				Leverages Chebyshev polynomials for lightweight cryptographic 
				authentication to reduce computational overhead in 5G IoV networks 
				during DDoS attacks & Analytical evaluation focused on cryptographic 
				efficiency & Significantly reduced computational cost while enhancing 
				system responsiveness under high traffic loads & A larger 
				message-signature tuple increases communication overhead, potentially 
				causing latency in dense networks \textbf{$\mid$\&$\mid$} It has not 
				been evaluated in real-time vehicular scenarios \\
				\hline
				
				``Optimized Feature Selection for DDoS Attack Recognition 
				and Mitigation in SD-VANETs'' \cite{TM3} & Hybrid (Statistical Analysis + ML) & Uses 
				optimized feature selection with LSTM models for effective mitigation, 
				focusing on reducing false positives & Emulated on real datasets with 
				statistical feature analysis & Achieved $94\%$ detection accuracy with 
				optimized mitigation strategies & It requires high computational 
				resources, challenging for low-powered vehicular devices \\
				\hline
				``DDoS Mitigation Based on Space-Time Flow Regularities 
				in IoV'' \cite{TM4} & Reinforcement Learning (RL)& Uses RL to adaptively adjust 
				mitigation strategies based on space-time flow patterns and disconnects 
				malicious connections through dynamic feature selection & Tested on the 
				Shenzhen taxicab dataset using the simulation tools 'ddosflowgen' and 
				'hping3 & Yielded high detection accuracy, but with increased time and 
				memory consumption & It needs high computational demand as it uses big 
				training data, limiting deployment in real-time IoV environments \\
				\hline
				``RSU-Based Online Intrusion Detection and Mitigation for 
				VANET'' \cite{TM5} & Statistical Anomaly Analysis &Uses roadside units (RSUs) to 
				filter and block malicious traffic in real-time and focuses on stealthy 
				DDoS mitigation using statistical methods to identify abnormal data 
				streams& Evaluated using real traffic datasets in a simulation 
				environment &Effectively reduced the impact of attacks on urban RSUs 
				with low false positive rates and delays & Less effective in areas with 
				low RSU density \textbf{$\mid$\&$\mid$} relying on centralized 
				infrastructure \\
				\hline
				\textbf{Our System in this Paper} & \textbf{Smart Quasi-Deterministic Forwarding 
					policy and Adaptive Attack Mitigation} & \textbf{Implements a smart 
					forwarding mechanism to regulate incoming traffic and uses adaptive 
					packet dropping to maintain network devices' performance during flood 
					attacks} & \textbf{Evaluated on a custom testbed with different flood 
					attack scenarios implemented using the MHDDoS public repository} & 
				\textbf{Significantly minimized the cost function that accounts for both 
					benign packet drops and the overhead associated with testing incoming 
					packets} & \textbf{Dropping benign packets can be a challenge during 
					long-time attacks \textbf{$\mid$\&$\mid$} Highly sensitive to parameter 
					tuning in complex IoV environments} \\
				\hline
			\end{tabular}%
		}
		\label{tab:mitigation_Table}
	\end{table*}
	
		\subsection{Organization of this Paper}
		
The previous Section \ref{Related} has reviewed the literature on AD and Mitigation for the IoT and the IoV, and Tables I and II compare recent state-of-the-art results with the work in the present paper.
	
	Our previous work \cite{BestPaper} presents experiments that show  the impact of Flood Attacks that often accompany cyberattacks on IoV Servers or Gateways, such as C-ITS Road Side Infrastructure Installations for city traffic control and dispatch \cite{Frotscher}, and  IoT Gateways for smart homes, buildings and factories  \cite{Augusto}.
		
		In Section \ref{Initial}, experimental results are shown concerning the system presented in Figure \ref{Zero-0}, which is comprised of a Gateway Server that supports an AD system and a set of connected devices that may forward benign or attack traffic to the Server.  In our experiments, a compromised Sensor sends UDP Flood attacks against the Server which receives network traffic from several Sensors. The measurements that are exhibited in this section, and which are not included in our earlier work \cite{BestPaper},  demonstrate the impact of the attack on the IoT Server, which creates congestion and {\bf disrupts the Server's operations} by impeding it from carrying out its important role of supporting the AD system that detects the attacks.
		
		We then review the use of a Smart  ``Quasi-Deterministic Forwarding Transmission Policy (QDTP)'' Forwarder (SQF) \cite{ICC22} in  Section \ref{QDTP}. In particular, we evaluate the performance of the architecture shown in Figure \ref{Forwarder} that includes the SQF, whose role is to shape the incoming traffic before it enters the Server. Our analysis and experimental results show that by choosing the SQF parameters judiciously, the undesirable effects of the Flood Attack on the Server are eliminated. However, the length of the input queue to the SQF increases significantly during the attack, and therefore a mitigation algorithm under the control of the AD system is needed.  
		
		Therefore, in Section \ref{AAM}, a novel attack mitigation algorithm using optimum AD and packet drop functions named ``Optimum AAM'' is proposed and its parameters are optimized. Optimum AAM is also extensively validated with experimental results and measurements. Its scalability is discussed in Section \ref{Scale}.

		Finally, Section \ref{Conc} presents our conclusions and suggestions for future work.

\section{The Experimental Test-Bed and its Behaviour Under a Flood Attack} \label{Initial}

\begin{figure}[t!]
			\centering
			\includegraphics[height=5.7cm,width=8.5cm]{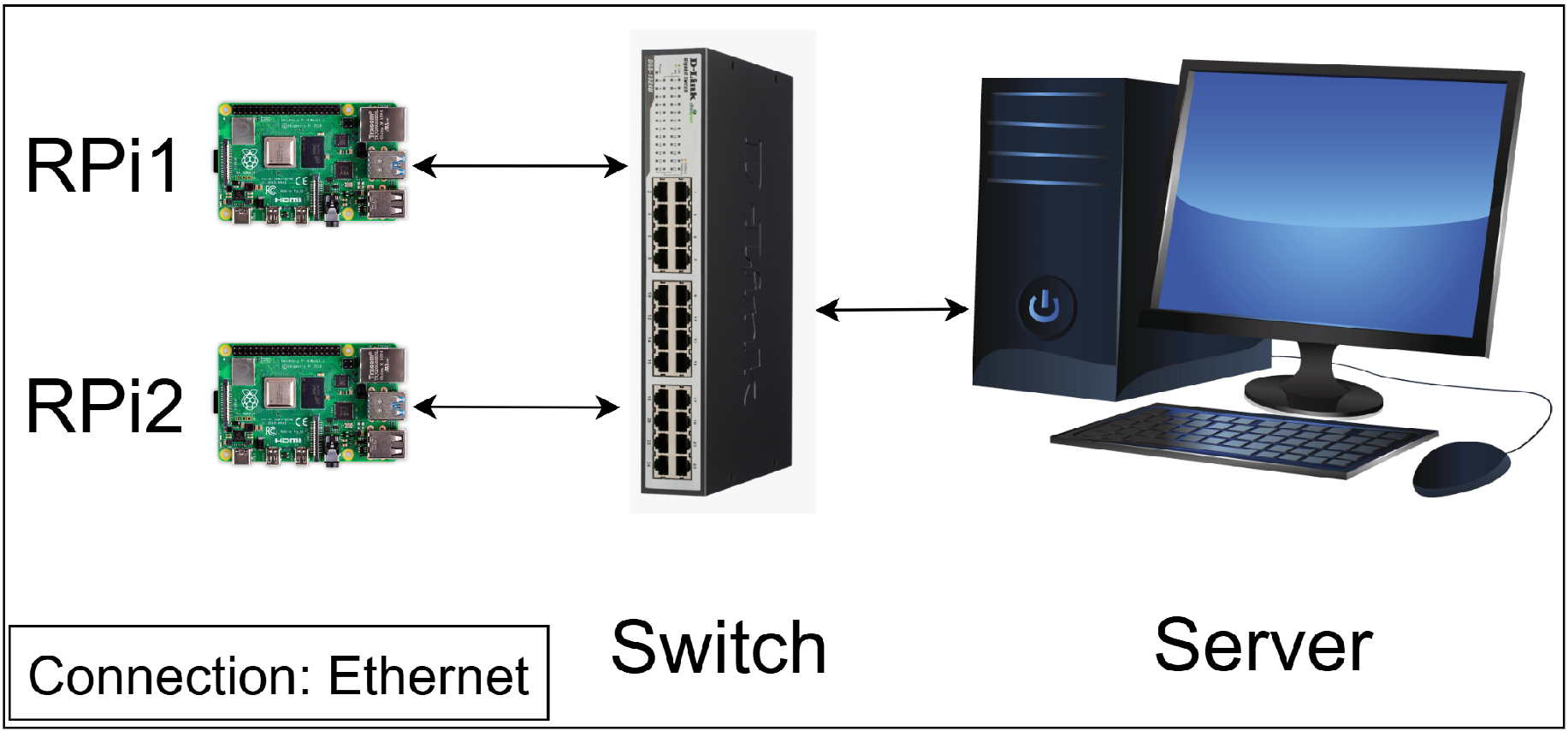}
			\caption{This figure shows the experimental test bed with several Raspberry Pi machines that emulate various devices, and are connected via Ethernet to the Server. The Raspberry Pis can send both normal and attack traffic to the  Server that acts as a Gateway for the IoV}
			\label{Zero-0}
		\end{figure}

The test-bed of Figure \ref{Zero-0} includes sensors emulated by 
Raspberry Pi $4$ Model B Rev $1.2$ (RPi$1$ and RPi$2$) computers. Each 
RPi is equipped with a $1.5 GHz$ ARM Cortex-$A72$ quad-core processor, 
$2GB$ $LPDDR4-3200$ SDRAM, and runs on Raspbian GNU/Linux $11$ 
(Bullseye), a Debian-based operating system. One RPi is programmed to 
send attack traffic packets randomly or in a predetermined manner, while 
the (uncompromised) RPi sends legitimate UDP packets that contain real 
data about the machines’ temperature periodically to the Server. The 
RPis have a network buffer size of $176$ kilobytes ($KB$) for both 
incoming and outgoing data.

\begin{figure}[t!]
			\centering
			\includegraphics[height=5.5cm,width=8.5cm]{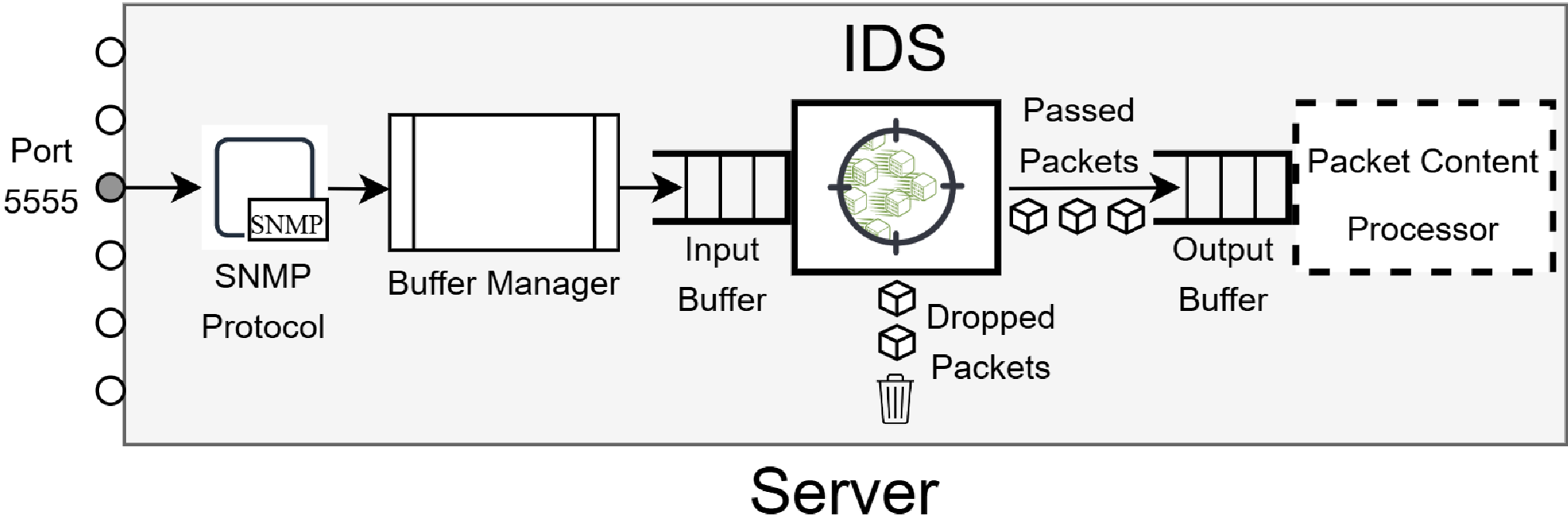}
			\caption{The software at the Gateway Server includes the manager for the SNMP network, the attack detection system (AD) \cite{CDIS}, as well as the software for processing the contents of incoming packets.}
			\label{Zero-1}
		\end{figure}

The Server is an Intel $8-Core i7-8705G$ processor running at $3.1 GHz$, 
with $16 GB$ of RAM, a $500 GB$ hard drive, and the Linux 
$5.15.0-60-66-Ubuntu SMP$ operating system. It communicates with low 
cost Raspberry Pi processors that emulate sensors through a shared Switch via Ethernet. Figure 
\ref{Zero-1} shows that the Server receives packets through the LAN 
using the UDP protocol and processes them using the Simple Network 
Management Protocol (SNMP) version $6.2.0-31$-generic. The Server's 
network interface card (NIC) supports a maximum speed of $1000 Mb/s$ in 
full duplex mode, with a network buffer size of $208 KB$ for incoming 
and outgoing data. Since UDP does not create connections and avoids 
ACKs, its lightweight nature is useful for communications with simple 
sensors and actuators \cite{UDP}.

The maximum size of the transmission unit (MTU) for both the Server 
and the RPi devices is set to $1.5 KB/Packet$, which optimizes packet size 
for efficient transmission. During a test with $1000$ successive packets, the 
measured latency ranged from $0.082 ms$ to
$0.514 ms$, with an average latency of $0.437 ms$, that indicates low 
latency across the network.

The Server also has an attack detection (AD) system detailed  in \cite{CDIS}. 	It is based on the Random Neural Network \cite{RNN1} extended with triggered movement of neuronal action potentials \cite{gelenbe1993g}, trained with DL \cite{gelenbe2017deep}.  
The dataset that is used is MHDDoS \cite{MHDDOS}, including up-to-date real-world DoS attacks and $56$ different attack emulators.  

\begin{figure}[t!]
			\hspace*{-27pt}
			\centering
			\includegraphics[height=6cm,width=10.2cm]{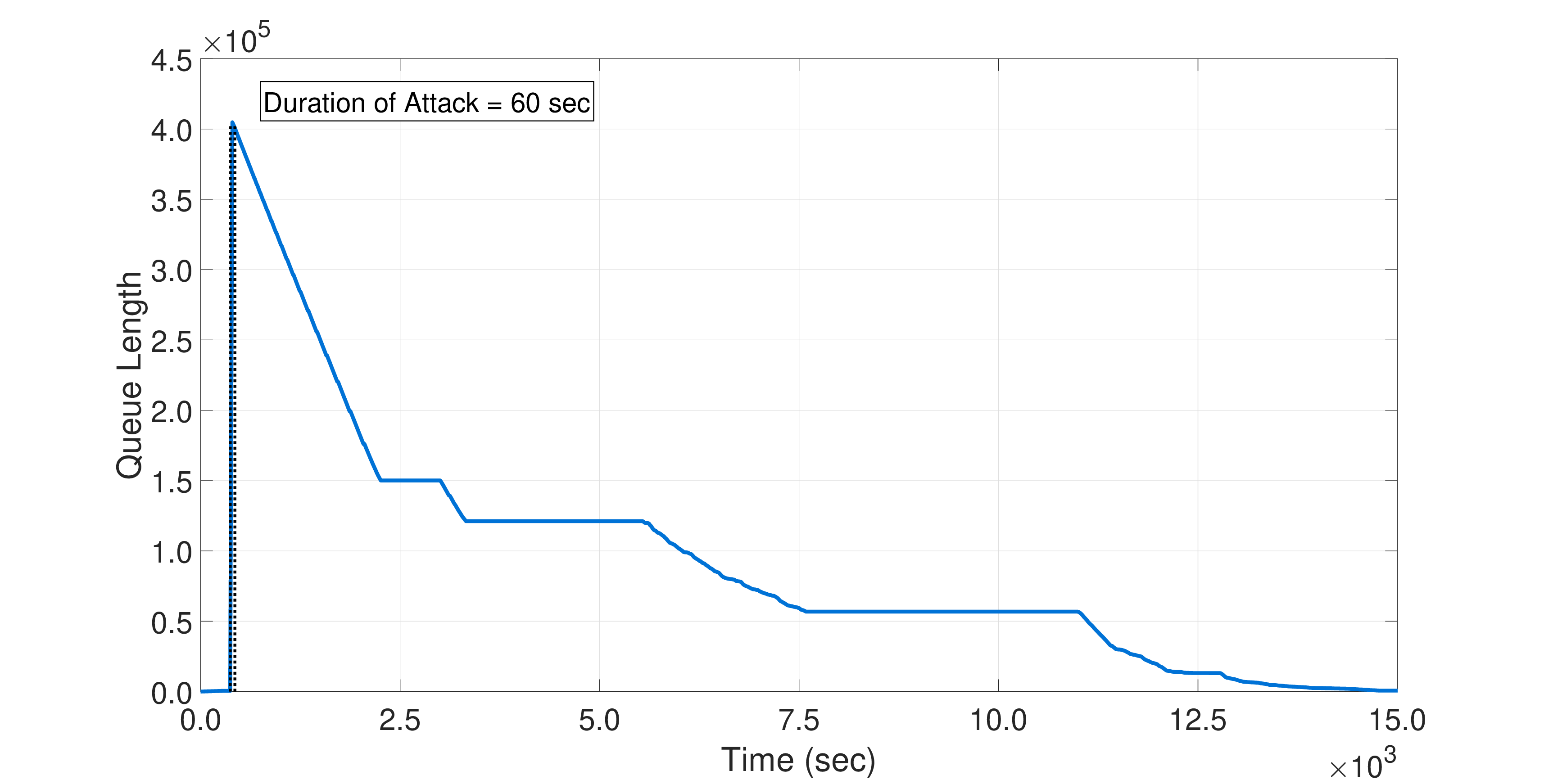}
			\caption{The queue length shown along the $y$-axis (number of packets) at the Server input prior to the AD,  shown as it varies over time ($x$-axis in seconds) at the Server input, prior to processing by the AD module, for a $60-seconds$ UDP Flood Attack that was launched by a Raspberry Pi in Figure \ref{Zero-0}. The queue length rises rapidly to $400,000$, and 
				the Server congestion then lasts far longer than the attack itself, i.e. up to several hours, because
				of Server paralysis which delays  AD processing, as seen in the  AD processing times of Figure \ref{fig:ProcessingTime2}.}
			\label{Zero}
		\end{figure}

As an example, Figure \ref{Zero} shows measurements 
from a Flood Attack that lasts for $60$ seconds (sec) and overwhelms
the capabilities of the 8-Core Server via an incoming flow of $400,000$ packets. The server becomes
paralyzed to the point where it is unable to operate the AD system. As a result, a huge packet queue forms at the Server's input, and it can only be progressively removed after a very long $300$ minute (min) delay. This also  impairs the Server's other normal activities, such as processing benign packets and removing the packets that are part of the attack.



	\subsection{The Server's Behaviour during Attack Detection}

\begin{figure}[h!]
	\hspace*{-27pt}
	\centering
	\includegraphics[height=6cm,width=10.2cm]{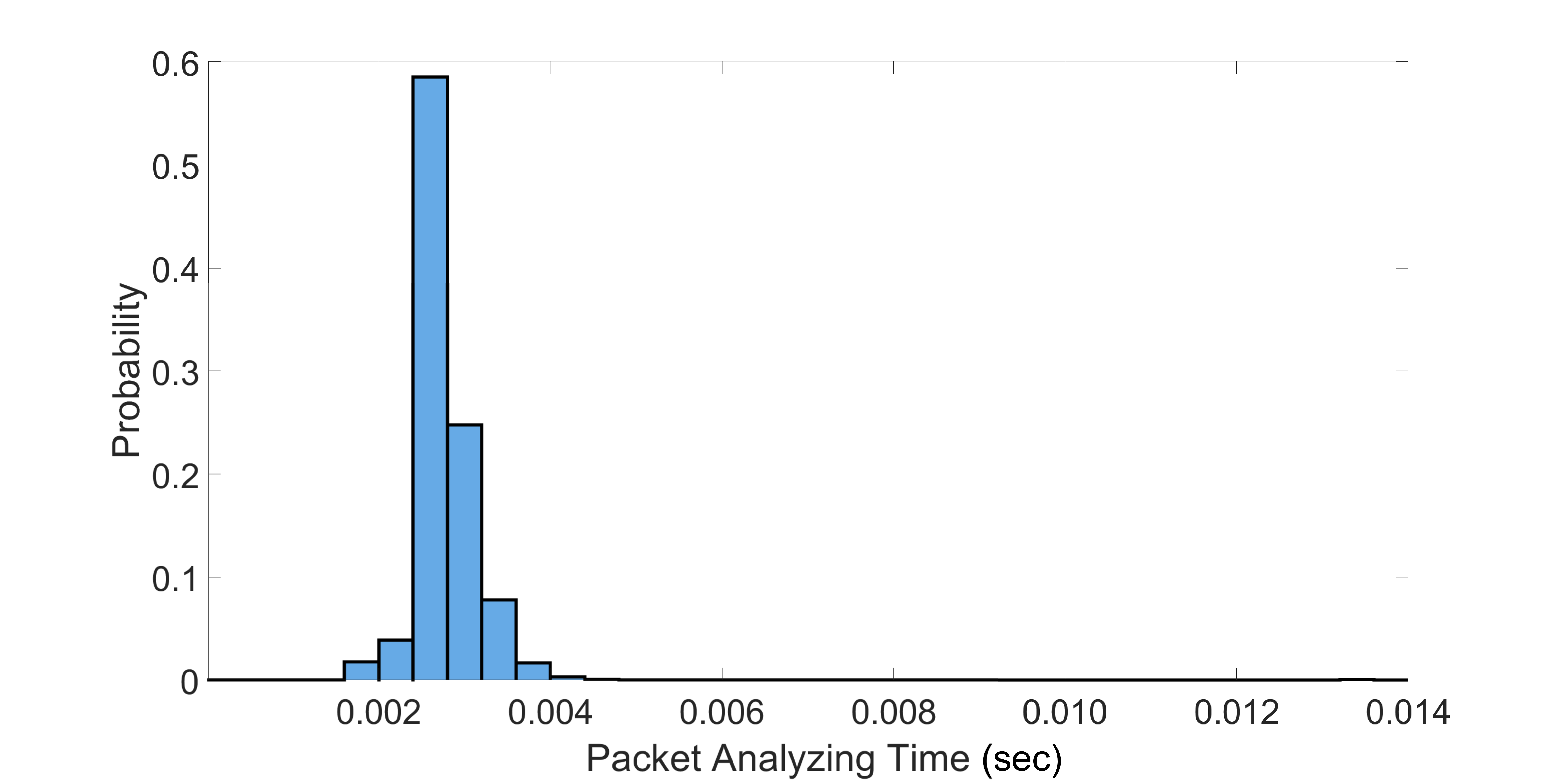}\\
	\hspace*{-27pt}
	\centering
	\includegraphics[height=6cm,width=10.2cm]{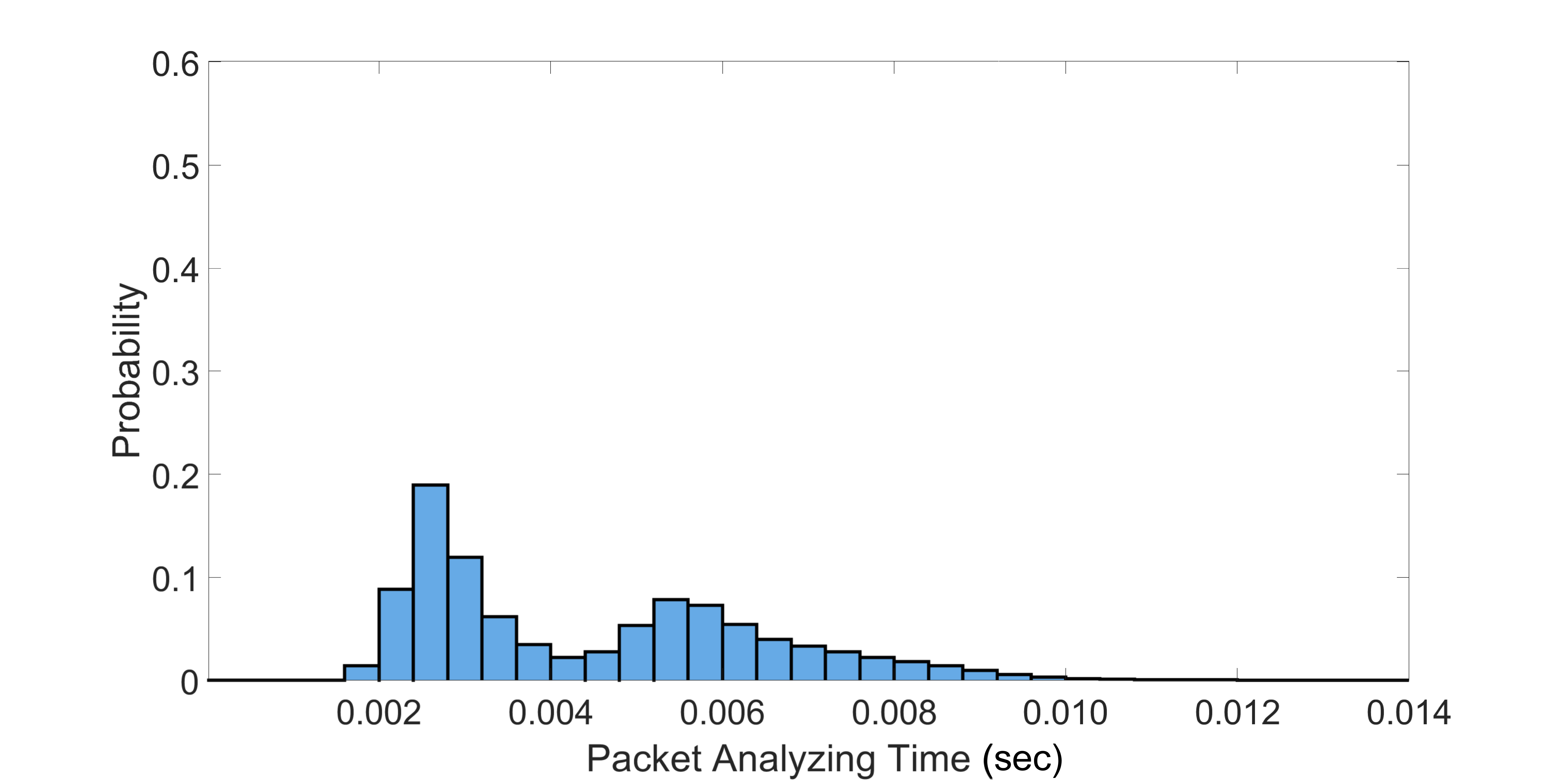}
	\caption{In the figure that is above, one can observe the histogram of AD processing time per packet, as measured without an attack. It shows the average processing time of $2.98$ ms (milliseconds), 
		with a variance of $0.0055~ms^2$. In the figure given below, an attack occurs and the AD packet processing time increases to the average value of  
		$4.82$ ms with  $0.51~ms^2$.}
	\label{TimePerPacket}	
\end{figure}

\begin{figure}[h!]
	\hspace*{-23pt}
	\centering
	\includegraphics[height=6cm,width=10.2cm]{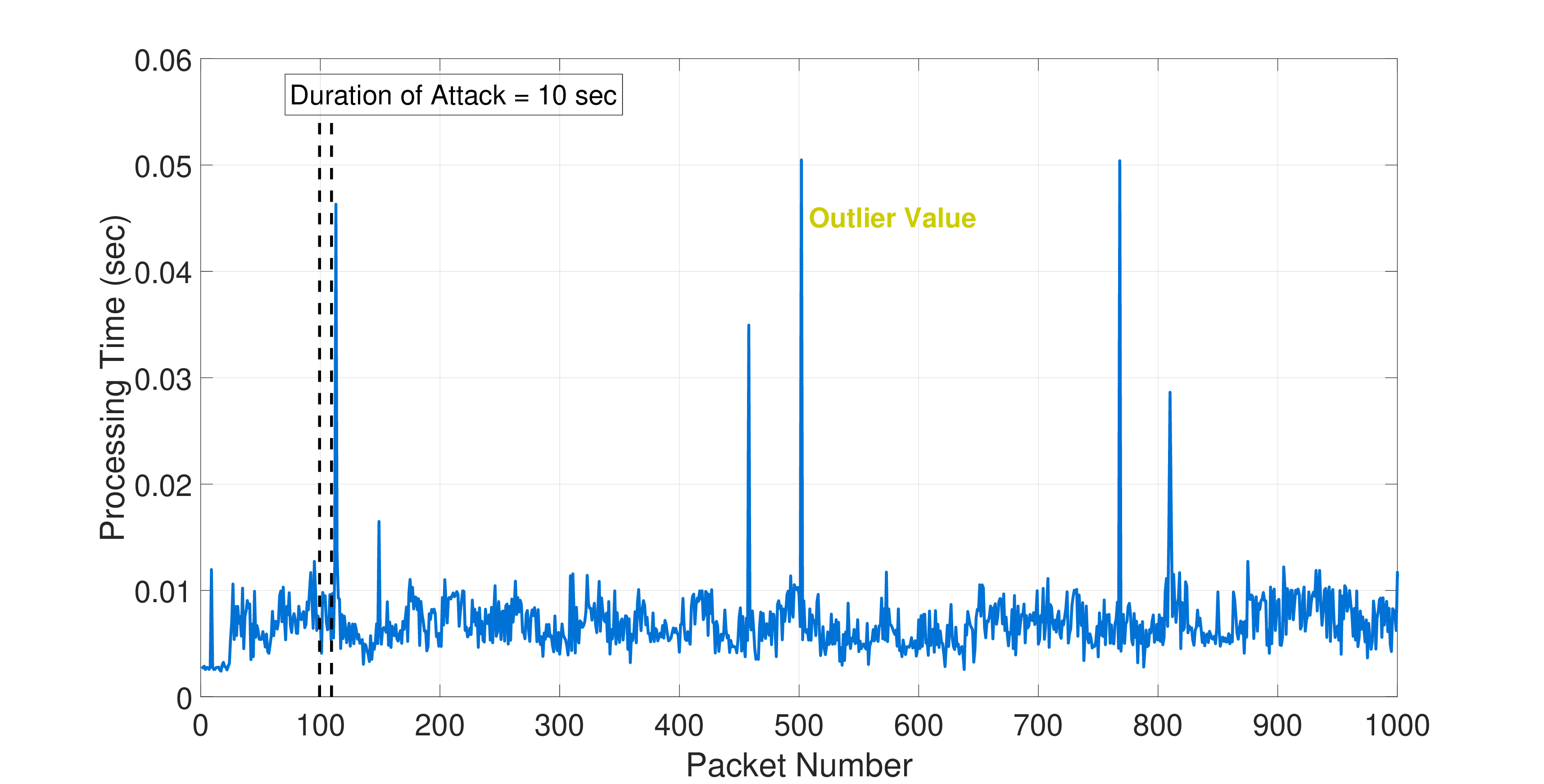}
	\caption{During a UDP Flood Attack, we show successive measurements for the AD packet processing time per packet when the QDTP Forwarder SQF is not used. The large outliers in processing time that are observed in Figure \ref{TimePerPacket}	
	(below) also confirm the measurements that are shown in this figure.}
	\label{fig:ProcessingTime2}
\end{figure}

Detailed measurements of the Server's AD processing times per packet both under normal conditions and during a UDP flood attack,  are reported in Figure  \ref{TimePerPacket} and  Figure \ref{fig:ProcessingTime2}. 

Figure \ref{TimePerPacket}  (above), indicates that the average  AD processing time per packet when there are {\bf no attacks} is $2.98$ milliseconds (ms), while (below) we see that when the Server is under a Flood Attack,  the
average processing time of the AD algorithm rises significantly to $4.82$ ms. When the Server is under attack, 
the AD processing time has very large outliers, as shown both in the histogram of Figure \ref{TimePerPacket}  (below), and in Figure \ref{fig:ProcessingTime2}. These outliers appear to be occasionally
$10^3$ times larger than the average. As a result, we conclude that the Flood Attack can significantly paralyze and slow down the Server's AD processing rate, as also illustrated in Figure \ref{Zero}. 

Thus, the {\bf AD algorithm itself is paralyzed} during a UDP Flood Attack, since the Server is overwhelmed by the SNMP protocol which processes the incoming packets.

\begin{figure}[t!]
	\centering
	\includegraphics[height=5.7cm,width=8.5cm]{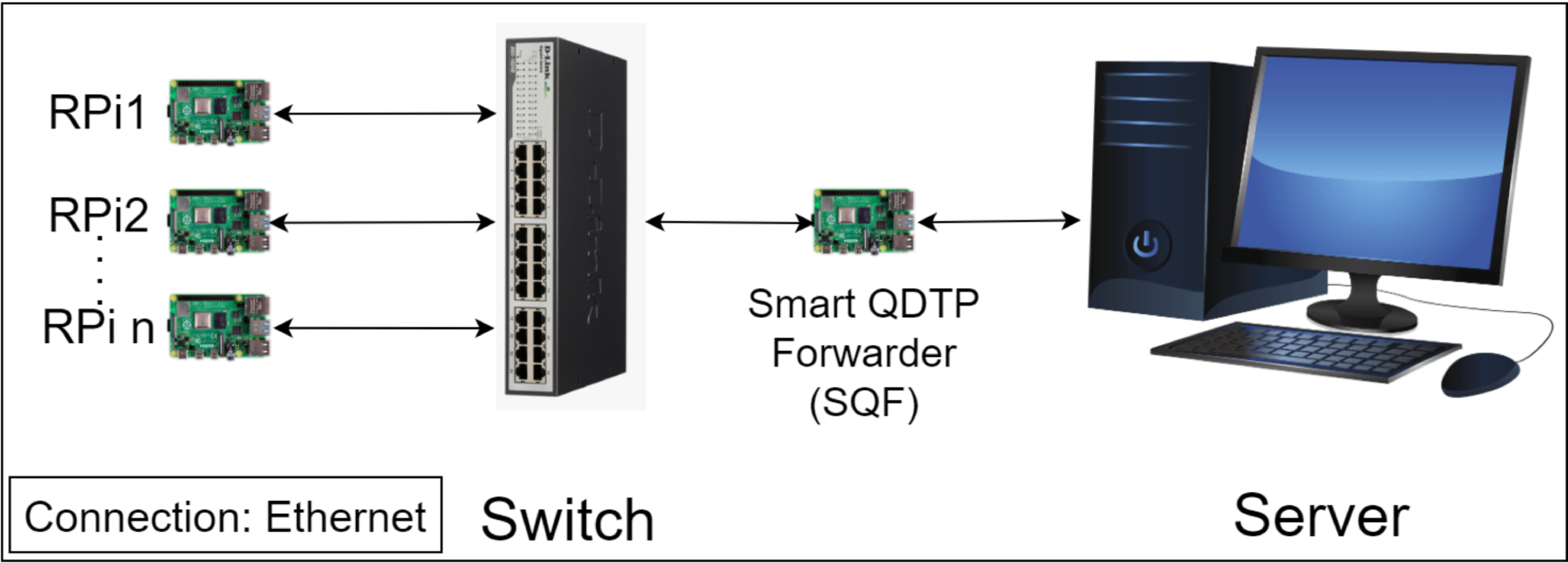}
	\caption{The figure shows the modified system architecture where a Smart QDTP Forwarder (SQF) is placed before the Server, and acts as a traffic
		shaping interface between the Ethernet LAN and the Server. The effect of the
		SQF is to eliminate the paralyzing effect of the
		packet flood at the Server, buffering packets within the SQF, and forwarding the packets, so that  AD processing and other work are conducted in a timely fashion.}
	\label{Forwarder}
\end{figure}

\begin{figure}[h!]
	\hspace*{-27pt}
	\centering
	\includegraphics[height=6cm,width=10.2cm]{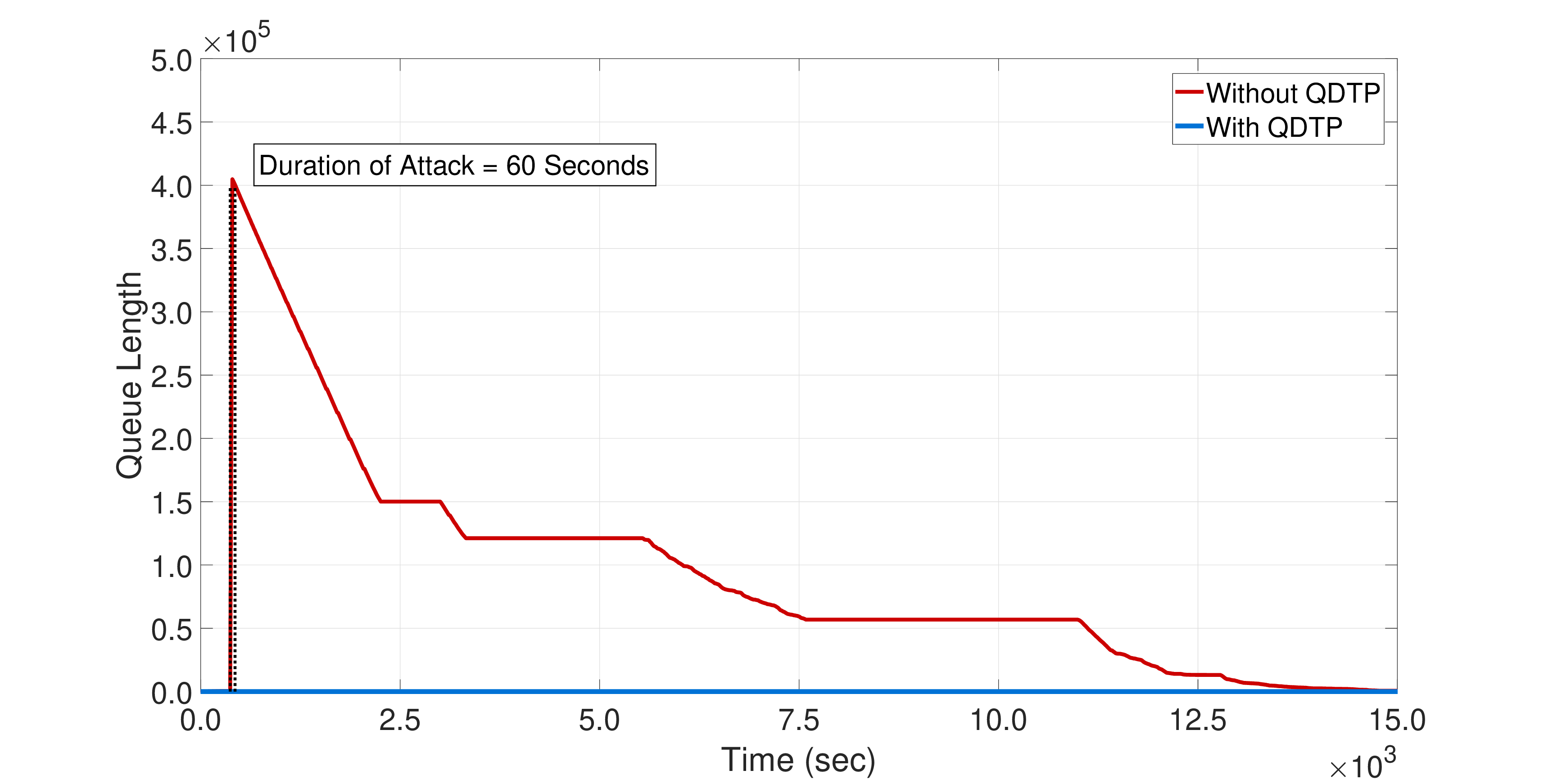}\\
	\hspace*{-27pt}
	\centering
	\includegraphics[height=6cm,width=10.2cm]{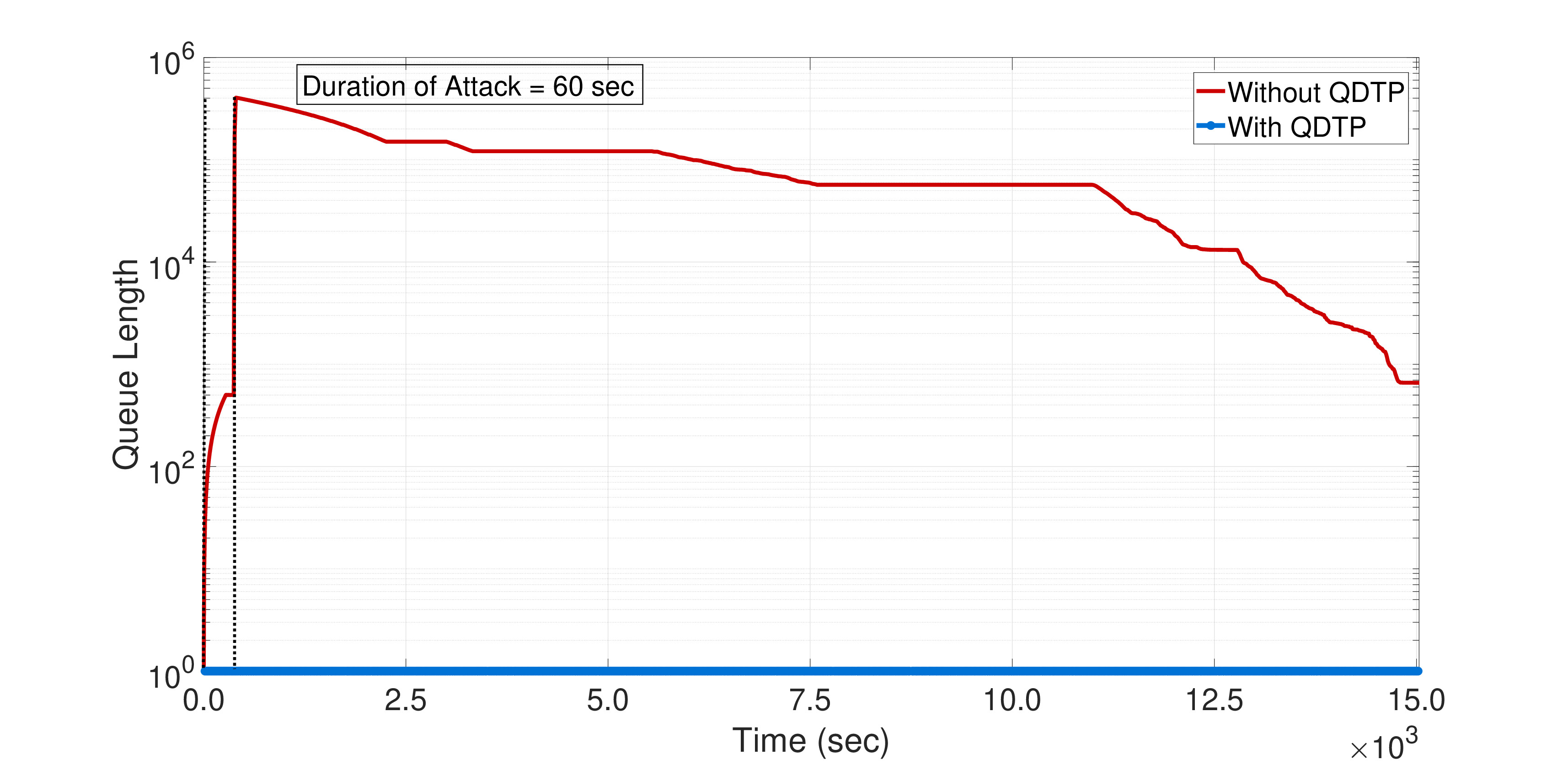}
	\caption{The Server queue length shown in the figure above, when a $60$-second UDP Flood Attack occurs, peaks at $400,000$ packets when the SQF is not used, and drops very slowly during some $15,000$ secs. In the figure that is below, we observe the 
		queue length in logarithmic scale  when the SQF is used with $D=3~ms$, versus the case
		shown in Red when the SQF is not used when the Flood Attack lasts $60$ secs. Since $D$ is close to the average of $T_n$ (which is $\approx 2.98$ ms when an attack does not occur, see Figure \ref{TimePerPacket}), small fluctuations of $T_n$ can cause the queue to build up moderately (in Blue).} 
	\label{QL-60s}
\end{figure}

\begin{figure}[h!]
	\hspace*{-21pt}
	\centering
	\includegraphics[height=6cm,width=10.2cm]{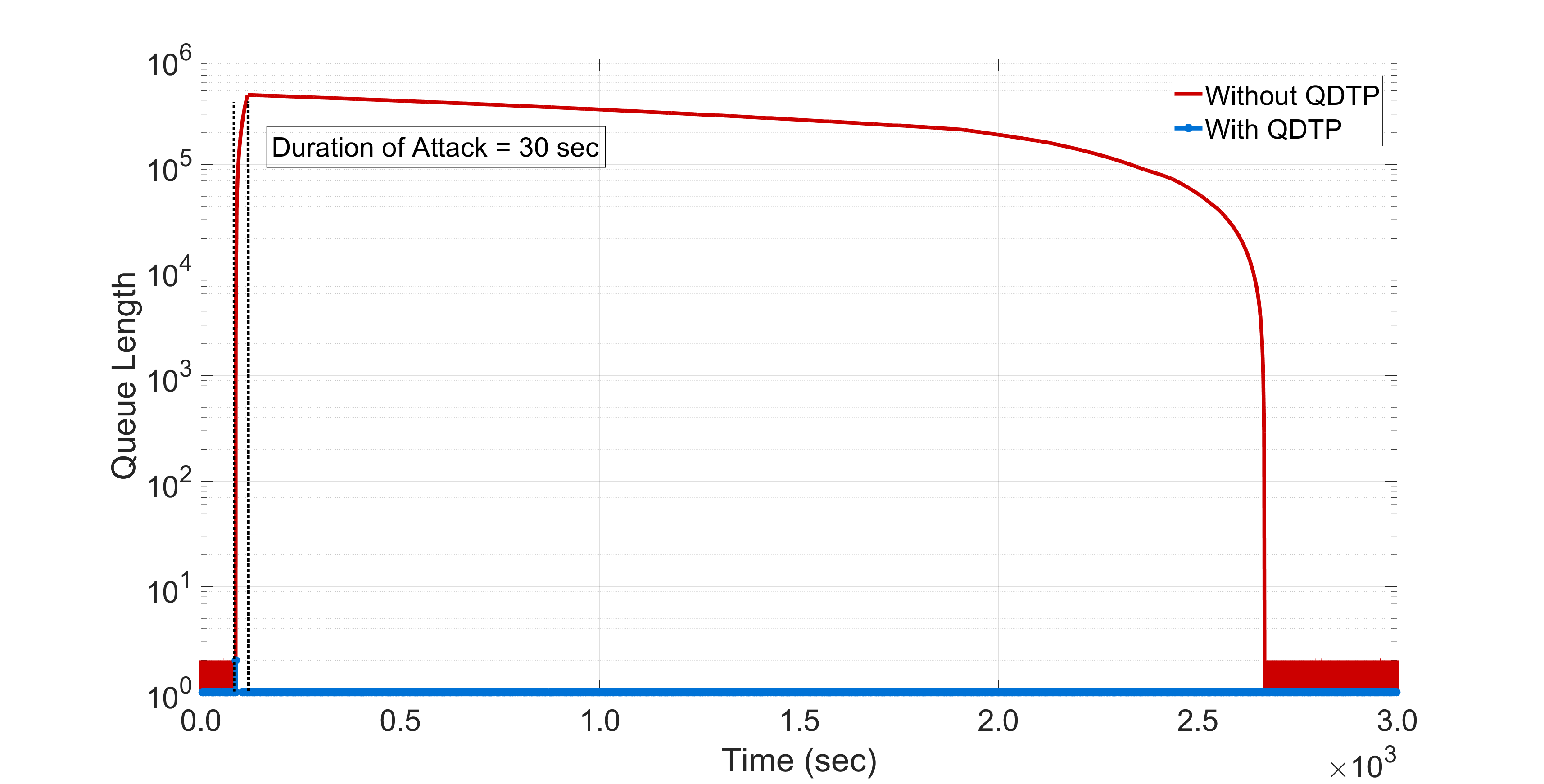}\\
	\hspace*{-21pt}
	\includegraphics[height=6cm,width=10.2cm]{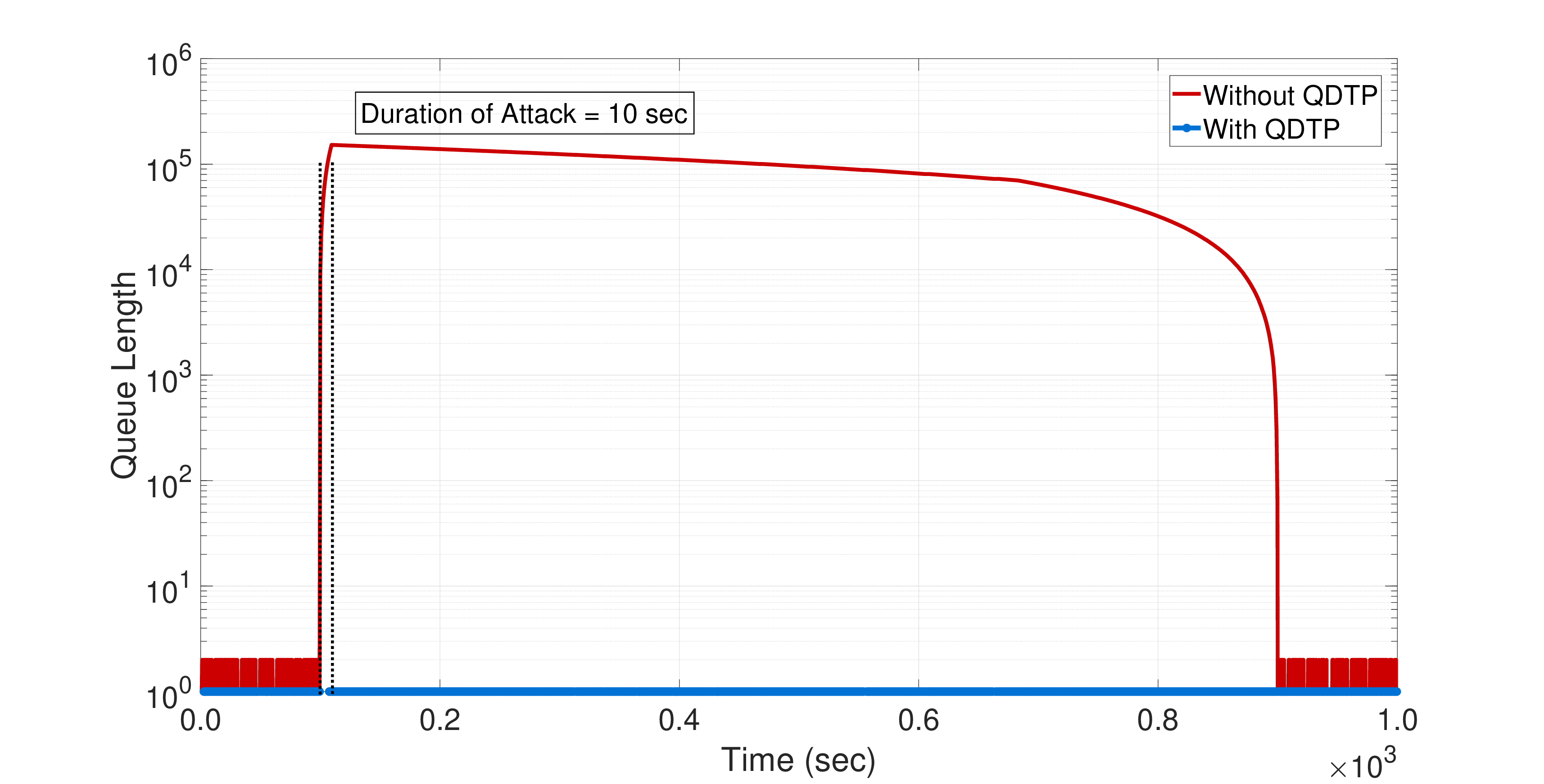}
	\caption{The Server queue length is measured, and represented logarithmically, for a UDP Flood Attack that targets the Server and lasts $30$ sec (above) or $10$ sec (below). In Red, we show the queue lengths  when the SQF is not used. The effect of using the SQF is shown in the Blue curves, demonstrating the effect of the SQF in reducing queue length during $30$ and $10$ sec attacks, with $D =3 ~ms$.}
	\label{QL-30s}
\end{figure}

\begin{figure}[t!]
	\hspace*{-27pt}
	\centering
	\includegraphics[height=6cm,width=10.2cm]{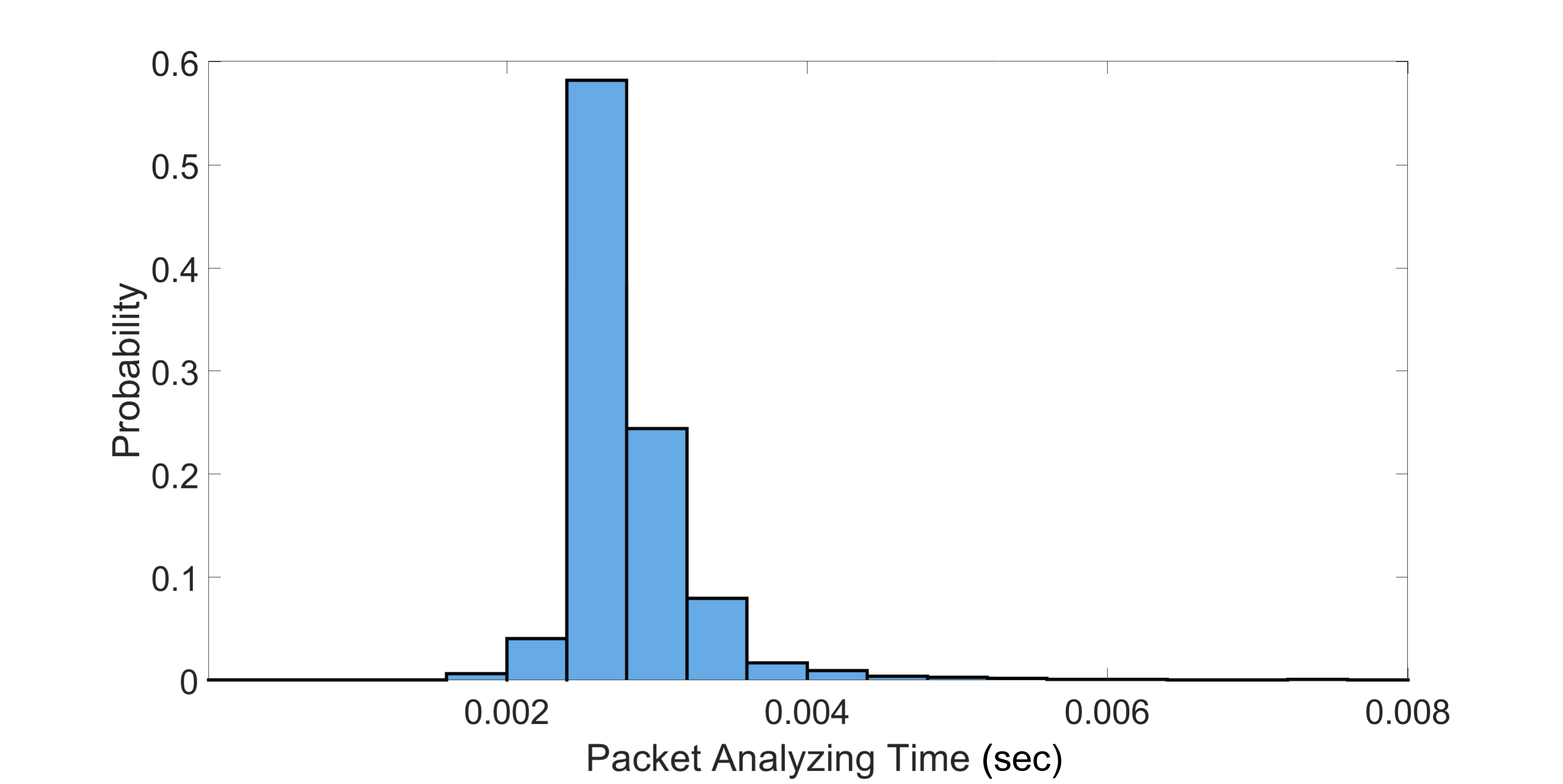}\\
	\hspace*{-27pt}
	\includegraphics[height=6cm,width=10.2cm]{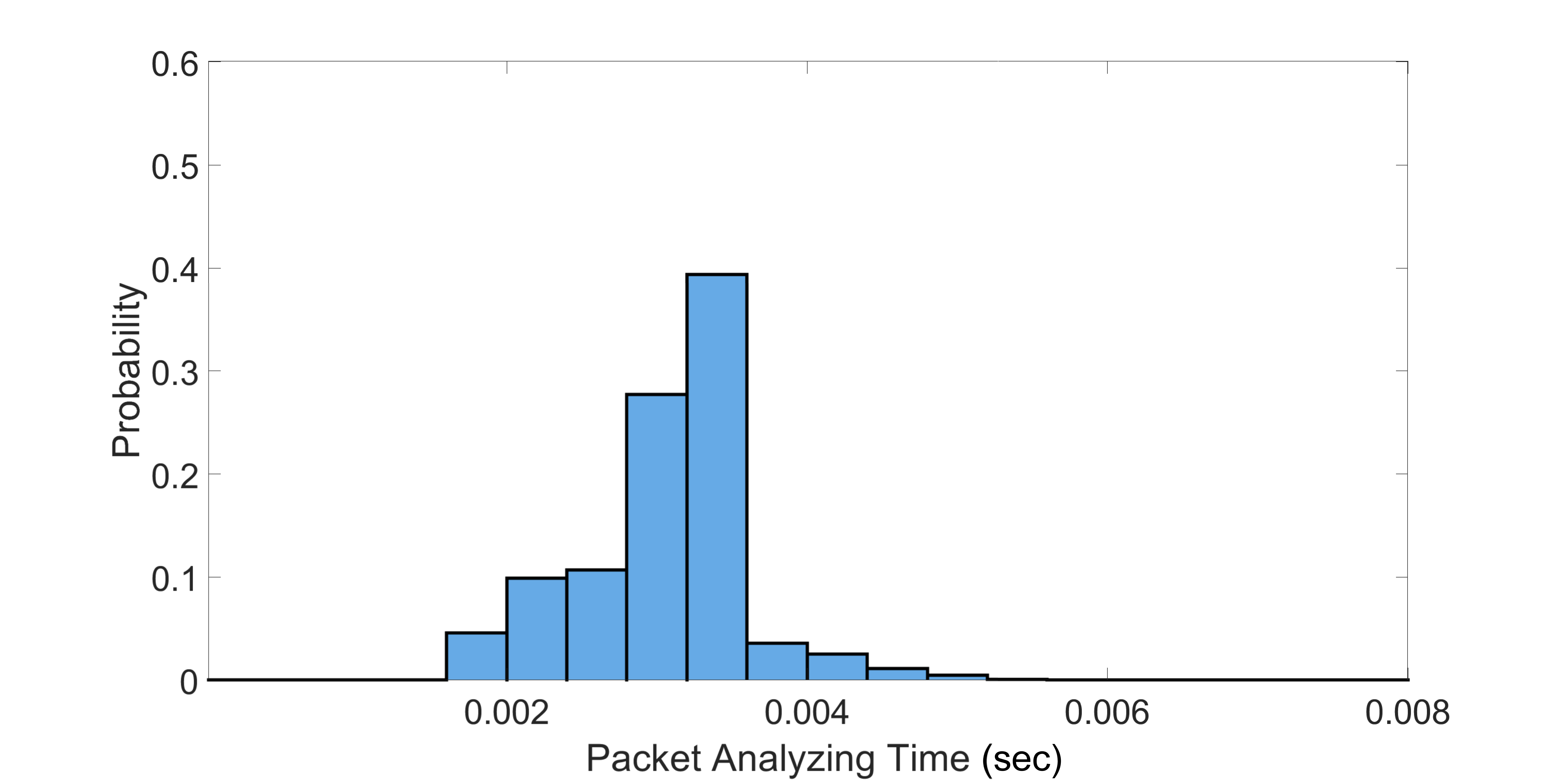}
	\caption{The AD Processing Time per packet  is shown at the Server when the SQF is used, and $D=2.7~ms$. The average AD packet processing time is
		$2.97~ms$ and its variance is $0.0041$ $sec^2$ when the attack occurs (above). The consequence of the Flood Attack (below) is to increase the AD average processing time of the AD per packet by just $10\%$ to $3.28~ms$, and a variance of $0.0023$ $sec^2$. This again demonstrates the effectiveness of the SQF.}
	\label{Fig4}
\end{figure}

\begin{figure}[t!]
	\hspace*{-21pt}
	\centering
	\includegraphics[height=6cm,width=10.2cm]{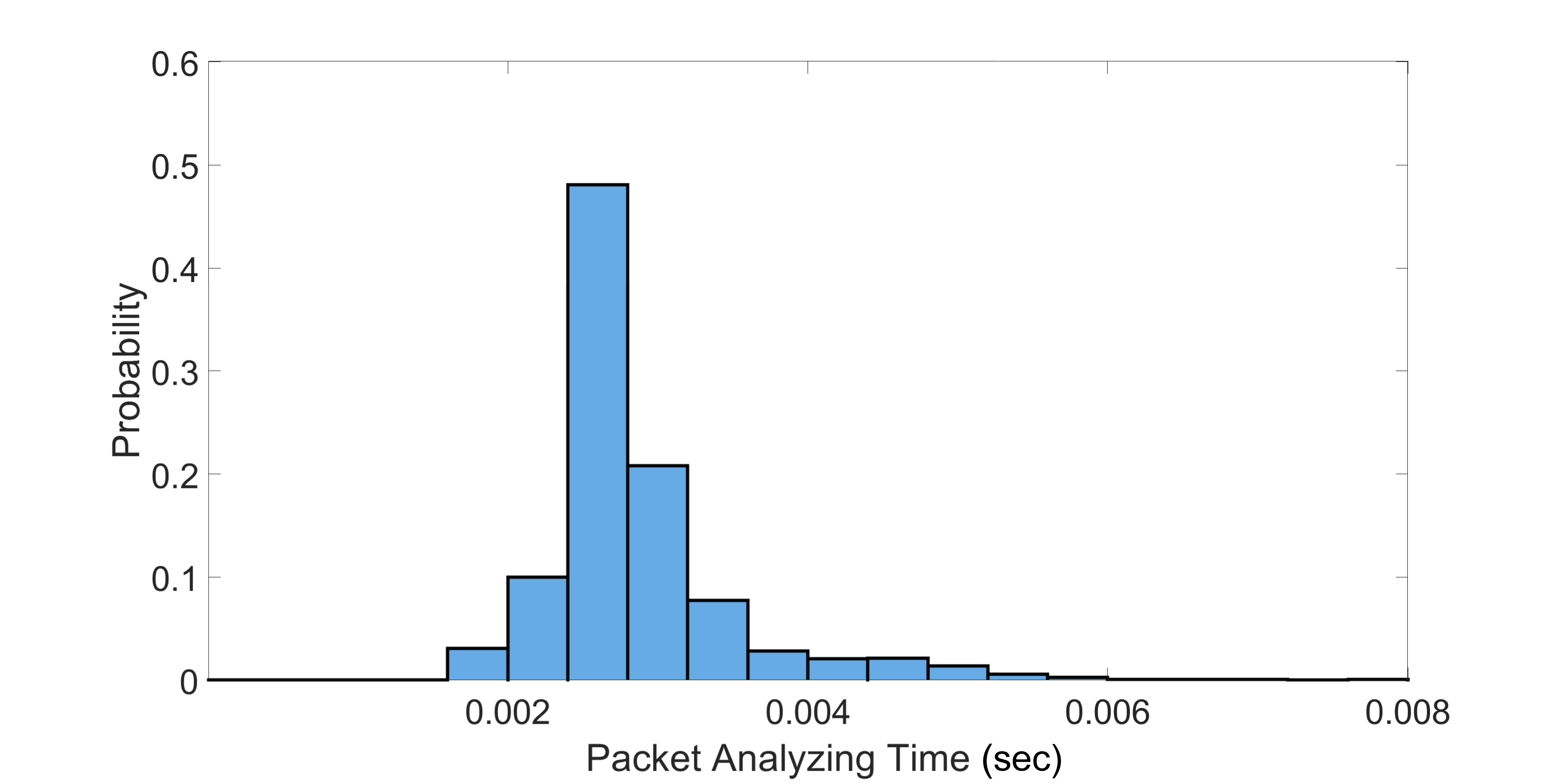}\\
	\hspace*{-21pt}
	\includegraphics[height=6cm,width=10.2cm]{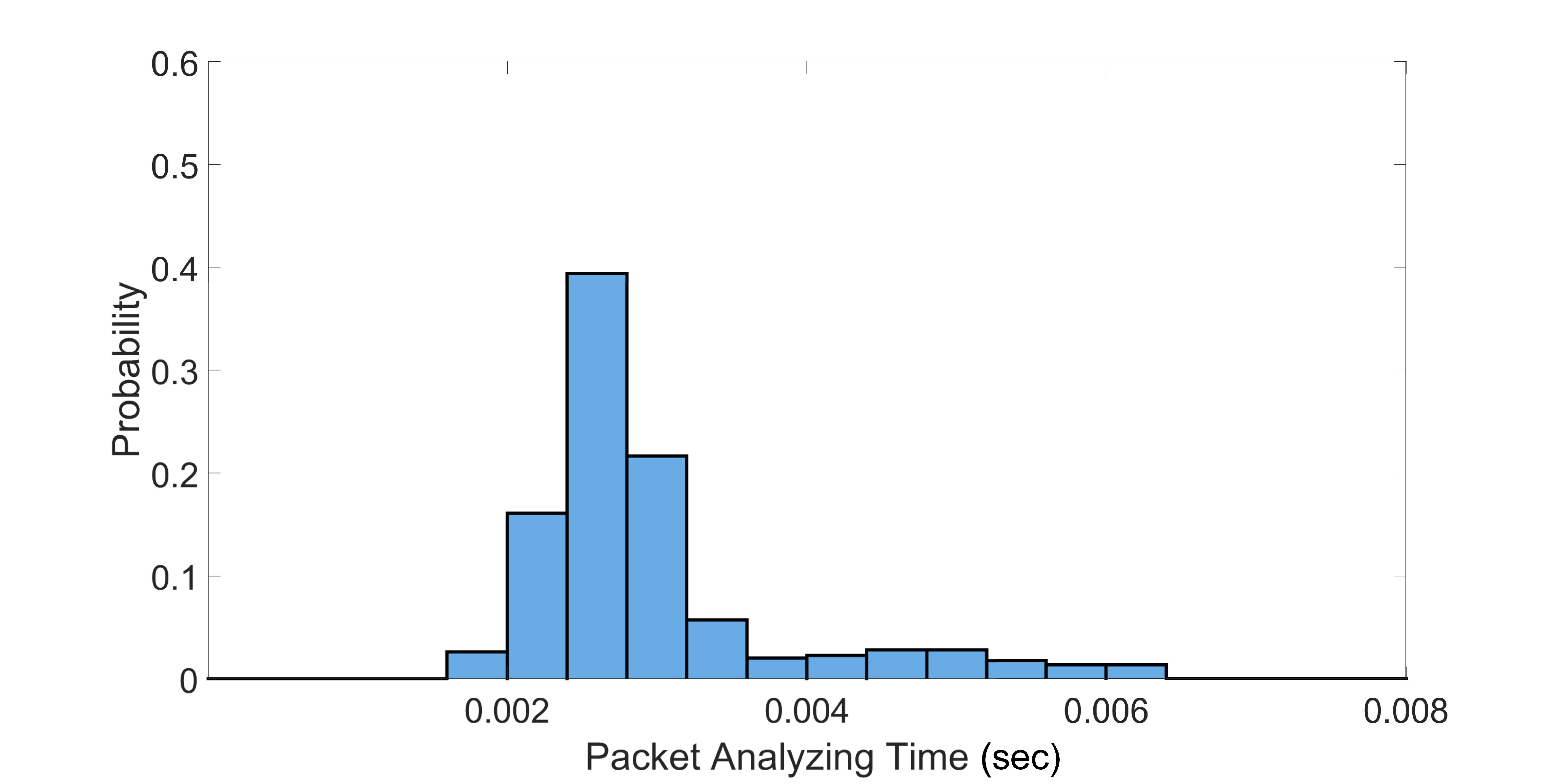}
	\caption{AD Processing Time at the Server with the SQF  and $D=3.20~ms$, resulting in an average AD processing time $T_n$ of
		$3.00~ms$ with a variance of $0.0036$ $sec^2$  when there is no attack (above). Under a Flood Attack (below),  the average AD processing time remains nearly the same at  $2.99~ms$ with a variance of $0.0067~sec^2$. This again shows that SQF is very effective in avoiding Server slowdown during an attack.}
	\label{Fig5}
\end{figure}
\begin{figure}[h!]
	\hspace*{-21pt}
	\centering
	\includegraphics[height=6cm,width=10.2cm]{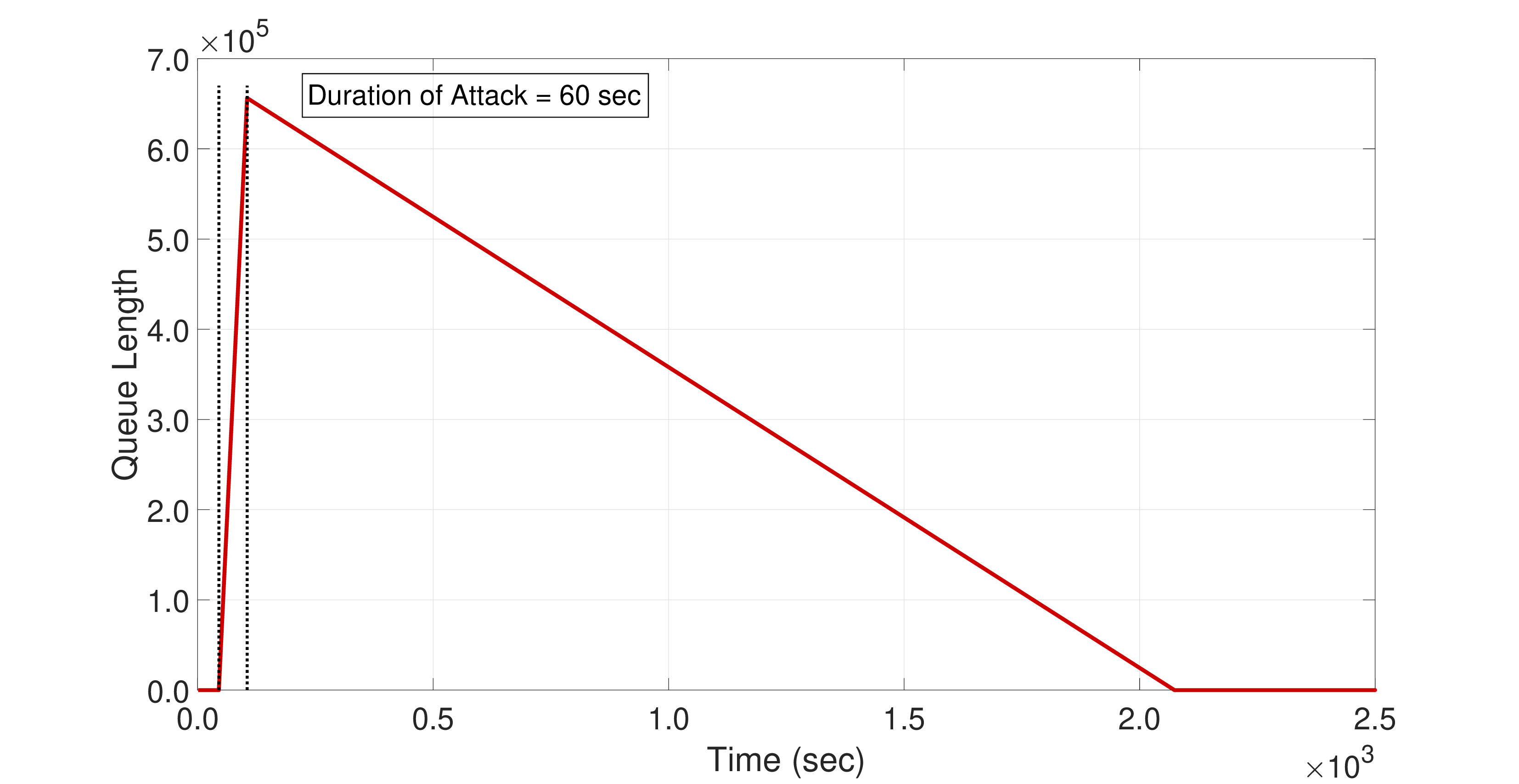}
	\caption{SQF queue length ($y$-axis in the number of packets) against time ($x$-axis in seconds) when a UDP Flood Attack lasts $60$ seconds, with $D =3~ms$, and without any mitigation action.}
	\label{fig:RPI-QL-30s-60s}
\end{figure}

		\subsection{Lindley's Equation when the SQF is not Used}

Let us now place ourselves in the context where the SQF module is {\bf not} being used (see Figure \ref{Zero-0}):
\begin{itemize}
	\item  Let $0=a_0\le a_1\le a_2,~ ...~$, be the successive packet arrival instants at the Server  through the Ethernet LAN from any of the Sensors connected to the LAN. We also define the interarrival time $A_{n+1}=a_{n+1}-a_n$. 
	\item Let $T_n$  denote the Server's AD processing time for the $n-th$ packet, and assume that
	the Server processes packets in First Come First Served (FCFS) order.
\end{itemize} 
Then the total waiting time $L_{n+1}$ of the $n+1$-th incoming packet, between the instant $a_n$ and the start of the AD processing time of the Server, is given by the well known Lindley's equation:
\begin{equation}
L_{n+1}=\max(0,L_n+T_n-A_{n+1}),~n\geq 0,~L_0=0~.\label{Lindley}
\end{equation}
Note that $L_0=0$ because the first incoming packet encounters an empty queue in front of the AD.
Note also that whenever we have $T_n>A_{n+1}$ then $L_{n+1}>L_n$, i.e. the waiting time increases.

During a Flood Attack, the values of $A_n$ and $T_n$ will be modified, as we see from
Figure \ref{Zero}, indicating that packet arrival rates have considerably increased so that the values of $A_n$ are much smaller, while Figure  \ref{TimePerPacket} (below) shows that the values of $T_n$ are also larger. However, the form of (\ref{Lindley}) does not change.

	\section{Effect of the Smart QDTP Forwarder (SQF)} \label{QDTP}

In Figure 	\ref{Forwarder}, we present our proposed modified architecture where the Server, whose role is to 
process incoming IoV packets -- including operating the AD module in order to detect attacks -- is ``protected'' by a Smart QDTP Forwarder (SQF) placed between the Ethernet output and the Server's input port.   The SQF shapes the incoming traffic to the Server with a  {\em Quasi-Deterministic Transmission Policy (QDTP)}  \cite{ICC22} that delays some of the packets it receives, by forwarding them to the Server at time $t_n \geq a_n$, where $a_n$ is the $n$-th packet's arrival instant to the SQF. 

We assume that the physical transmission time from the SQF to the Server, and the network protocol service time inside the Server, are negligible with respect to the Server's AD processing duration $T_n$.  This is consistent with our measurements with $1000$ packets, that revealed a latency between $0.082 ms$ and
$0.514 ms$, with an average value of $0.437 ms$, which is less than  $15\%$ of the value of $T_n$ (which is around $3~ms$).

 Thus, when the $n-th$ packet is transmitted by the SQF, it is assumed that it instantly arrives at the queue at the Server's input and  forwarded for AD processing. Therefore, the instant $t_n$ when SQF forwards the $n$-th packet to the Server is initiated as $t_0=a_0$. For a constant parameter $D$ that needs to be selected, it is given by the following recursive expression:
\begin{eqnarray}
&&t_{n+1}=\max(t_n+D,a_{n+1}),~n\geq 0,\label{eq1}\\
&&hence:~t_{n+1}-t_n\geq D~. \label{eq2}
\end{eqnarray}
The total delay $Q_n$ experienced by the $n$-th packet due to the SQF, elapsing from the arrival of the $n$-th packet to the SQF at $a_n$, until its arrival to
the AD at the Server at $t_n$, is then:
\begin{eqnarray}
&&Q_0=t_0-a_0=0,\label{Q0}\\
&&Q_{n+1}=t_{n+1}-a_{n+1},\nonumber\\
&&=\max(t_n+D,a_{n+1})-a_{n+1},\nonumber\\
&&=0,~if~t_n+D\leq a_{n+1},~and\nonumber\\
&&=t_n+D-a_{n+1},~otherwise.
\end{eqnarray}
Since $t_n=Q_n+a_n$, we obtain the recursive expression:
\begin{eqnarray}
&&Q_{n+1}=\max(0,t_n+D-a_{n+1}),\nonumber\\
&&=\max(0,Q_n+D-A_{n+1}),~n\geq 0,\label{Q}
\end{eqnarray}
which is also an instance of Lindley's equation (\ref{Lindley}).

On the other hand, the Server's AD module also acts as an FCFS queue and we can exploit
Lindley's equation again to compute $W_n,~n\geq 0$ the waiting time of the $n$-th packet that arrives at the Server to be processed for attack detection, which is:
\begin{eqnarray}
W_{n+1}&=&\max(0,W_n+T_n-(t_{n+1}-t_n)),~W_0=0,\nonumber\\
&\leq &W_n+T_n-(t_n-t_{n+1}), \label{Server2}
\end{eqnarray}
since the $n$-th packet's AD service time is $T_n$ and the $n+1$-th interarrival interval to the
Server's AD queue is $t_{n+1}-t_n$. Therefore, from (\ref{eq2}) and (\ref{Server2}) 
we have:
\begin{equation} 
W_{n+1}\leq W_n + T_n - D, \label{Server3}
\end{equation}
and we obtain the  following key result which tells us how to choose $D$:

\medskip
\noindent{\bf Result 1.} If $D$ in the SQF is chosen to satisfy the inequality 
$D>T_n$ for any $n\geq 0$, then $W_n$, the packet waiting time  at the Server, will  be
$W_n=0,~\forall ~n\geq 0$.

\subsection{Experiments that Illustrate the Value of Result 1}

Since the data in Figure \ref{TimePerPacket} (above) shows the
histogram of the AD processing time per packet $T_n$ with average value $2.98$ ms when there is no attack, we set $D=3 ~$ms, just above that average as indicated by  Result 1.

The experimental results in Figure \ref{QL-60s} illustrate  the case {\bf without SQF} (above) and {\bf with SQF} (below) during a $60~sec$ UDP Flood Attack. Note that the figure above represents the Server queue length varying over time, without the SQF. The figure below shows the Server queue length with a logarithmic scale, and compares the cases without SQF (in red) and with SQF (in blue) for the  Server queue length that varies over time. Since $D=3$~ms is very close to the
average of $T_n$, the fluctuations in the values of $T_n$ cause the buildup of a short queue 
of a few packets, as seen  in the blue plot shown below. 

Figure 	\ref{QL-30s} shows the results of four experiments where we measure the queue length at the Server when a UDP Flood Attack lasts $30$ (above) and $10$ (below) seconds, without (red) and with (blue) the SQF. Without the SQF, the Server's AD processing time  increases significantly.
In the $30~sec$ attack, approximately $470,000$ packets are received at the Server and without the SQF it takes $44.45$ minutes for the Server to return to normal process them, while in the $10~sec$ attack $153,667$ packets are received,
and it takes the Server roughly $15$ minutes to process them. Note that in these curves, it takes
some $99$ seconds for the compromised RPi to launch the attack.

Figure \ref{Fig4}  ~shows that  when we use the SQF based system  with $D=2.7~ms$, which is 
smaller than the value recommended by  Result 1,  when there is no attack, this choice of $D$ 
has very little effect. However, when a UDP Flood Attack occurs, the Server's AD processing is somewhat slowed down, and the average value of $T_n$ increases by roughly 10\%.

On the other hand, Figure \ref{Fig5} confirms {\bf Result 1}  since it shows that if we take $D=3.2~ms$, which guarantees that $D>T_n$ most of the time, then the measured average value of $T_n$ remains at around $3~ms$ showing that it has not been slowed down by the attack's overload effect. Of course the same is seen when no attack occurs.

\subsubsection{SQF Queue Buildup and Attack Mitigation} \label{Mitigate}

During a Flood Attack,  packets accumulate in the SQF input queue. The QDTP algorithm forwards them to the Server with $D=3~ms$, and we observe that the  Server does not experience any significant slowdown with regard to AD, as shown in Figure \ref{fig:RPI-QL-30s-60s}. We note the sudden queue length increase to over $600,000$ packets, followed by a very slow queue length decrease during more than $2,000$ secs, for a Flood Attack that only lasts $60$ secs.

This motivates us to develop the novel, and hitherto unpublished, Adaptive Attack Mitigation (AAM) method  that is discussed in Section \ref{AAM}.

\section{Adaptive Attack Mitigation (AAM)} \label{AAM}

In the previous Section, we observed that the SQF allows the AD to operate effectively during an attack by limiting the input packet rate.  However, the SQF does not stop the huge build-up of attack packets at the input of the SQF,  where the queue length increases to a very high value, as seen in
Figure \ref{fig:RPI-QL-30s-60s}. Without an effective mitigation scheme, these packets will have to be processed long after the attack itself may have stopped, even though they are largely attack packets that are of no use to the system. Thus, in this section, we propose a novel Adaptive Attack Mitigation (AAM) scheme, which will:
\begin{itemize}
	\item Reduce the amount of AD testing that is carried out during an attack, and hence reduce the computational overhead at the Server,
	\item Drop attack packets and reduce the effect of traffic congestion and overload both during and after an attack,
	\item Stop the DROP process in a timely fashion when the attack ends, to avoid the excessive loss of benign packets.
\end{itemize}

\subsection{Attack Detection AD} \label{AD}

\begin{figure}[h!]
		\hspace*{-10pt}
	\centering
	\includegraphics[height=6cm,width=9.8cm]{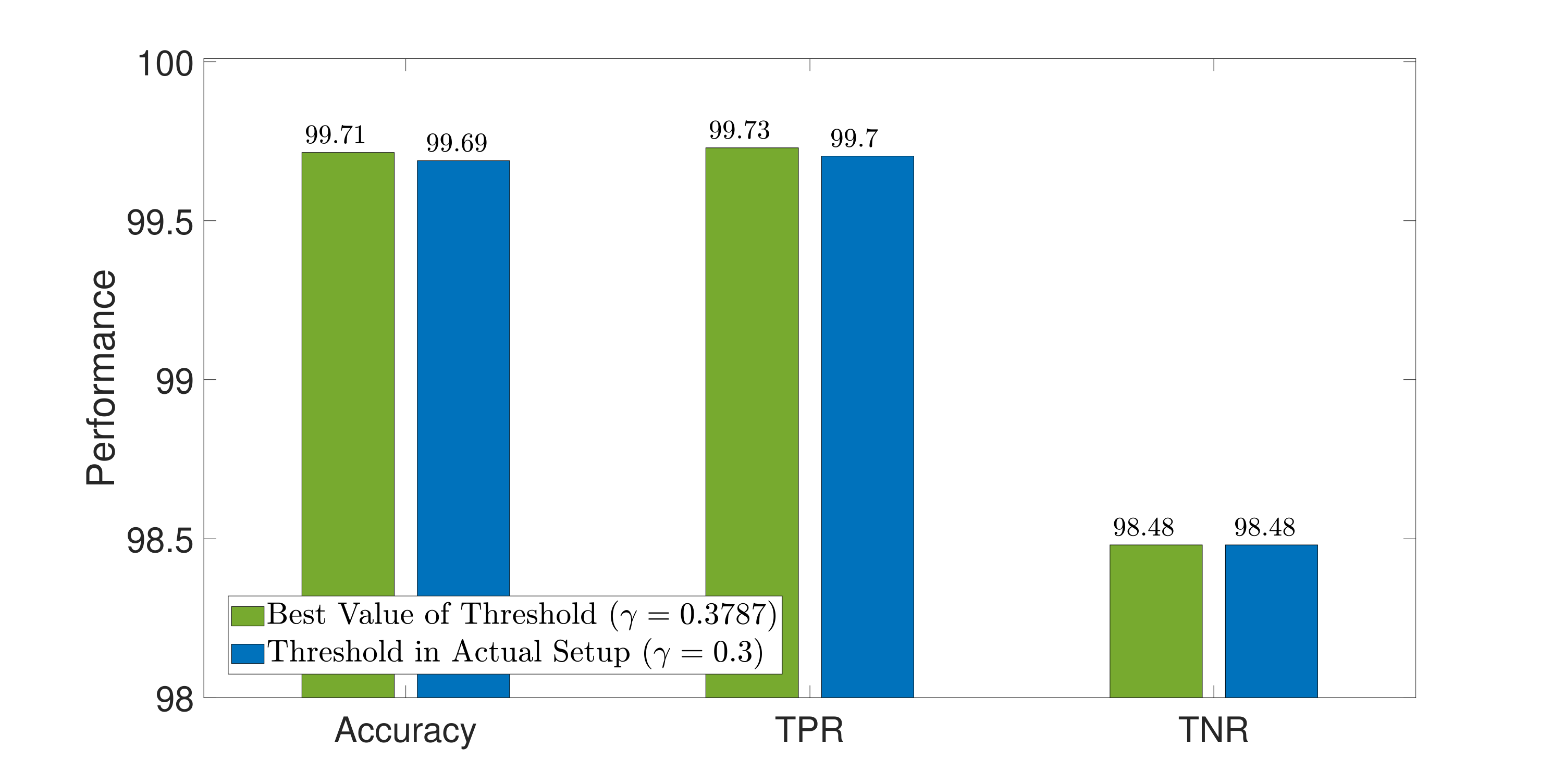}
	\caption{Accuracy of the AADRNN attack detector, that was evaluated on the test-bed of Figure 1.}
	\label{Accuracy}
\end{figure}

Recall (again) that the AAM uses an AD based on the Auto-Associative Deep Learning Random Neural Network described in \cite{brun2018deep}. It was evaluated with the  well known Kitsune attack dataset \cite{kitsune_paper,kitsune_dataset}, yielding  the highly accurate  results shown in Figure \ref{Accuracy}.  

The AD reaches a decision based on a Window of $W>0$ successive packets that are tested sequentially. If the AD system detects that a majority of the $W$ packets are of ``ATTACK'' type, then it concludes that an attack is occurring. Otherwise, it will return a NO-ATTACK decision.  If a NO-ATTACK is detected, then the AD system will proceed to test the subsequent $W$ packets in the same way. Note that $W$  is chosen empirically to be the smallest value of the window that provides high accuracy, and is typically set in the range of $8$ to $10$ packets. 

When the  AD detects an ATTACK, since the AD is substantially slowed down by the large packet backlog that accumulates  during the attack, the novel AAM technique that we propose, drops the preceding $m+W$ packets, and then skips ahead $m>0$ packets in the incoming packet stream, reaching a subsequent Testing Window of $W$ packets. Thus, the packet backlog is reduced, and the processing slowdown for AD is also reduced, since each individual packet is only tested by the AD when the system is {\bf not} under attack. 

Thus, AAM has the advantage of early attack detection, rapid dropping of blocks of attacking packets, and reducing the AD processing time during an attack by carrying out AD in successive $m$-packet intervals 
using a $W$-packet window. The AAM algorithm is detailed below: 
\begin{enumerate}
	\item Initialize $i\leftarrow 1~and~j\leftarrow 1$.
	\item Test the $W$ packets $i,i+1,~...~,i+W-1$ for AD.
	\item If the output of the AD is {\bf ATTACK then}:\\
	- {\bf DROP the PACKETS $j,j+1,~...~,i+W-1$},\\
	- {\bf Update $j\leftarrow i+W~and~i\leftarrow i+W-1+m$}
	\item If the output of the AD is {\bf NO-ATTACK then}:\\
	-  {\bf FORWARD Packets $j,j+1,~...~,i+W-1$ to the SERVER},\\  
	- {\bf Update $j\leftarrow i+W~and~i\leftarrow i+W$},
	\item {\bf Go To 2) }
\end{enumerate}
The summarized pseudocode and the detailed pseudocode for the AAM are given below as {\bf Algorithm 1} and {\bf Algorithm 2}. 

\begin{algorithm}
	\caption{Summarized Pseudocode for the AAM}
	\begin{algorithmic}[1]
	\State Initialize $i \gets 1$, $j \gets 1$
		\State Let $X$ be the total number of packets received during the attack
		\State Let $f$ be the fraction of attack packets
		\State Let $\alpha$ and $\beta$ be weighting factors for the cost function
		\State Compute the initial optimal value of $m$:
		\[
		m^* \gets \sqrt{2\frac{\beta}{\alpha} \cdot W(E[X] - W)} - W
		\]
		\While{there are incoming packets to process}
		\State Test the $W$ (window) packets: $i, i + 1, \ldots, i + W - 1$ for Attack Detection
		\If{Attack Detection output is \textbf{ATTACK}}
		\State Drop the packets $j, j + 1, \ldots, i + W - 1$
		\State Update $j \gets i + W$ and $i \gets i + W -1 + m^*$
		\State Update the total number of dropped packets.
		\State Calculate the expected value of the dropped benign packets cost.
		\State Update the total average cost of AAM.
		\Else \Comment{If Attack Detection output is \textbf{NO-ATTACK}}
		\State Forward the packets $j, j + 1, \ldots, i + W - 1$ to the server
		\State Update $j \gets i + W$ and $i \gets i + W$
		\State Update the total average cost of AAM.
		\EndIf
		\State Recalculate $m^*$ periodically as needed.
		\EndWhile
	\end{algorithmic}
\end{algorithm}
\begin{algorithm}
	\caption{Detailed Pseudocode for the AAM}
	\begin{algorithmic}[1]
		\State Initialize $i \gets 1$, $j \gets 1$
		\State Let $X$ be the total number of packets received during the attack
		\State Let $f$ be the fraction of attack packets
		\State Let $\alpha$ and $\beta$ be weighting factors for the cost function
		\State Compute the initial optimal value of $m$:
		\[
		m^* \gets \sqrt{2 \cdot \frac{\beta}{\alpha} W(E[X] - W)} - W
		\]
		\While{there are incoming packets to process}
		\State Test the $W$ (window) packets: $i, i + 1, \ldots, i + W - 1$ for Attack Detection
		\If{Attack Detection output is \textbf{ATTACK}}
		\State Drop the packets $j, j + 1, \ldots, i + W - 1$
		\State Update $j \gets i + W$ and $i \gets i + W -1 + m^*$
		\State Update the total number of dropped packets:
		\[
		\delta \gets N  (m + W)
		\]
		\State Calculate the expected value of the reprocessing cost dropped benign packets:
		\[
		E[K] \approx W \cdot \tau \cdot \left[\frac{(1-f)E[X]}{W} -\frac{1}{2} + \frac{m}{2W}\right]
		\]
		\State Update the total average cost of AAM:
		\[
		C(AAM) = \alpha \cdot E[K] + \beta \cdot E[\Omega]
		\]
		where
		\[
		E[\Omega] \approx \tau \cdot W \cdot \left[\frac{E[X] - W}{m^* + W} + \frac{1}{2}\right]
		\]
		\Else \Comment{If Attack Detection output is \textbf{NO-ATTACK}}
		\State Forward the packets $j, j + 1, \ldots, i + W - 1$ to the server
		\State Update $j \gets i + W$ and $i \gets i + W$
		\State Update the total average cost of AAM.
		\EndIf
		\State Recalculate $m^*$ periodically as needed.
		\EndWhile
	\end{algorithmic}
\end{algorithm}

		\subsection{Analysis and Optimization of AAM}

During a Flood Attack, a large fraction, say $0<f\leq 1$, of the incoming traffic will be part of the attack,and  the complementary fraction $(1-f)$ will be benign traffic coming from various sources that also send traffic to the IoV Gateway. Thus, the AAM may drop useful benign packets,  as well as attack packets. 

In addition, even though the AAM reduces the number of packets that are actually tested for AD during an attack, the proposed AAM still creates computational overhead during an attack, because  it tests $W$ packets after each $m$ packet interval.   Interestingly, the proportion of benign packets which are dropped will increase as $m$ increases, but at the same time the overhead also decreases as $m$ increases. Thus, we will now compute the optimum value of $m$. In the next subsection, we will validate these analytical results through some experiments.

Let $X$ denote the total number of packets received at the  Gateway during an attack, including a fraction $f$ of attack packets. Since $X$ cannot be known in advance, we treat it here as a random variable. Note that we will denote the expected value of the random variable $Y$ by $E[Y]$. 

During the attack, there will be a first inevitable $W$-packet AD window when the attack is first detected. This first testing window where the AD says ``ATTACK'' will contain a majority of attack packets among the $W$ packets. Since $W$ is very small compared to $X$, we can assume that the attack begins at the beginning of the first testing window where the AD reports an attack.

The end of the attack is signalled to the AAM by the AD during the first detection window in which a majority of the $W$ packets are not attack packets, which  indicates that  the current attack has ended. Thus, in addition to the first attack detection window, the total number of additional attack detection windows that are used during an attack is given by:
\begin{eqnarray}
N&=&\lceil \frac{X-W}{m+W}\rceil, ~and~\nonumber\\
E[N]´&\approx&\frac{E[X]-W}{m+W} + \frac{1}{2},\label{approx}
\end{eqnarray}
where the expression (\ref{approx}) is actually a mathematically proven \cite{ApproxFormula} first order approximation.
In particular, it applies to all probability density functions for the random variable $X$  which are a convex sum of Erlang densities, commonly used to approximate discrete histograms.

If the AAM is used in conjunction with the SQF that includes the QDTP traffic shaping policy, then Figure 10 shows that the AD average processing time per packet remains constant at roughly $3~msec$ which we will call $ \tau$, which varies with the speed of the processor at the Server. For each detection window of $W$ packets, the Server overhead is then $N\tau W$, so that  the resulting overhead for attack detection when an attack occurs is:
\begin{equation}
\Omega =N\tau W,~with~E[\Omega] \approx  \tau W[\frac{E[X]-W}{m+W} + \frac{1}{2}].
\end{equation}
On the other hand, the total number of dropped packets due to the attack is:
\begin{eqnarray}
\delta&=&W + N\times m + (N-1)\times W=N(m+W),\nonumber\\
E[\delta] &\approx&E[X]+\frac{1}{2}(m-W),
\end{eqnarray}
since the packets in the first window that says "ATTACK'', as well as the following $m$ packets are dropped. This is repeated a total of $N$ times, while those in the packets in the last window that say ``NO-ATTACK'' are not dropped. 

However, the part $X$ of $\delta$ includes attack and benign packets. Let the fraction of benign packets in $X$ be $(1-f)$. Since we know that the last part $\delta~-X$ is only composed of benign packets, the reprocessing time $K$ of {\bf the benign packets which the AAM drops} may create additional overhead, since it is likely that they will be sent again to the  Server after some length of time, and the Server will have to process them again using the AD. Hence:
\begin{eqnarray}
&&K= \tau W \lceil\frac{(1-f)X + \delta -X}{W}\rceil,~and~\nonumber\\
&&E[K]\approx \tau W [\frac{(1-f)E[X]}{W} -1+ \frac{m+W}{2W}]~,\nonumber\\
&\approx& \tau W [\frac{(1-f)E[X]}{W} -\frac{1}{2} + \frac{m}{2W}]~,\label{dropped}
\end{eqnarray}
and the total average cost of $AAM$ written as $C(AAM)$, assuming a weighting of $\alpha,~\beta>0$ for the two average cost terms $E[K]$ and $E[\Omega]$, respectively, becomes:
\begin{eqnarray}
&&C(AAM)=\alpha E[K] + \beta E[\Omega ] \nonumber\\
&&\approx W\tau \big[~\alpha [\frac{(1-f)E[X]}{W} - \frac{1}{2}+\frac{1}{2}\frac{m}{W}]\nonumber\\
&&+ \beta [\frac{E[X]-W}{m+W} + \frac{1}{2}]\big]. \label{cost}
\end{eqnarray}
Thus, taking the derivative of the right-hand-side of (\ref{cost}) with respect to $m$, and setting it to zero, we can see that the {\bf total average cost} $C(AAM)$ is 
{\bf approximately minimized by setting} $m$ to the value $m^*$:
\begin{equation}
m^* \approx \sqrt{~2\frac{\beta}{\alpha}W[~E[X]-W~]~}~-W,\label{opt-m}
\end{equation}
where $\frac{\beta}{\alpha}$ is the relative importance of the two terms in the cost function; $\beta<<\alpha$ when the benign packets that were dropped during an attack arrive (for the second time) at the Server while it is busy processing other incoming benign packets.

Interestingly, we see that $m^*$ does {\bf not depend on} $f$ and $\tau$. Furthermore, as the average number of packets $E[X]$ received by the Server during the attack increases, the optimum value $m^*$ also increases in proportion to the square root of $E[X]$.

\subsection{Scalability} \label{Scale}

We now discuss the scalability of the Optimum AAM Algorithm. The scalability will be discussed in the context of a Gateway Server with multiple network ports, since the AD and the Optimum AAM are designed to protect a {\bf single} Gateway Server. Let us first recall that the choice of the parameter $W$ is discussed in Section \ref{AD}, and that it is chosen empirically as the smallest value that provides sufficiently high accuracy for the AD algorithm, typically around $W\approx 8~to~10$ consecutive packets.	

For a single Gateway Server with $P$ network ports, we can use the Optimum AAM in several ways: (a) we may have one AD and one AAM for each port, or (b) one common AD and then one AAM per port, or (c) one AD for each port and a single common AAM, or (d) one AD and one single AAM for all the ports. In all these cases the value of $W$ will be the same.

Notice that the total processing time of all the ADs is proportional to the total number of packets that they process; thus the AD processing time is not affected by the configurations (a)-(d), but simply by the total traffic entering the Gateway from all ports; if all ports carry an equal amount of traffic, then the total AD overhead processing time will increase linearly with $P$. However, in the cases (b) and (d) where there is a single AD for all ports,in order to determine which port is being attacked,  the AD will have to group incoming packets into W-windows of packets with the same incoming port number, which requires additional computation. Therefore, with regard to the AD processing time, the solutions (a) and (c) are more efficient than (b) and (d).  

Similarly, the AAM is only activated when an attack is detected, so that the total number of AAM activations which cause processing overhead is the same in all cases and depends on the total number of attacks that aim the Gateway Server, rather than on the number of ports. However, one may argue that more ports will attract more attacks. On the other hand, having a single Optimum AAM unit, as in cases (c) and (d), has two shortcomings: (i) when multiple attacks occur within short intervals, a backlog of processing requests may accumulate in front of the single AAM software unit, leading to delays in the mitigation process, and (ii) the single AAM will have to deal with multiple distinct values of $m^*$, leading to sub-optimal decisions. Thus we recommend the use of (a) for the best performance and responsiveness to attacks.

\subsection{Optimization Measurements of the AAM Scheme}

\begin{figure}[h!]
	\hspace*{-21pt}
	\centering
	\includegraphics[height=6cm,width=9.5cm]{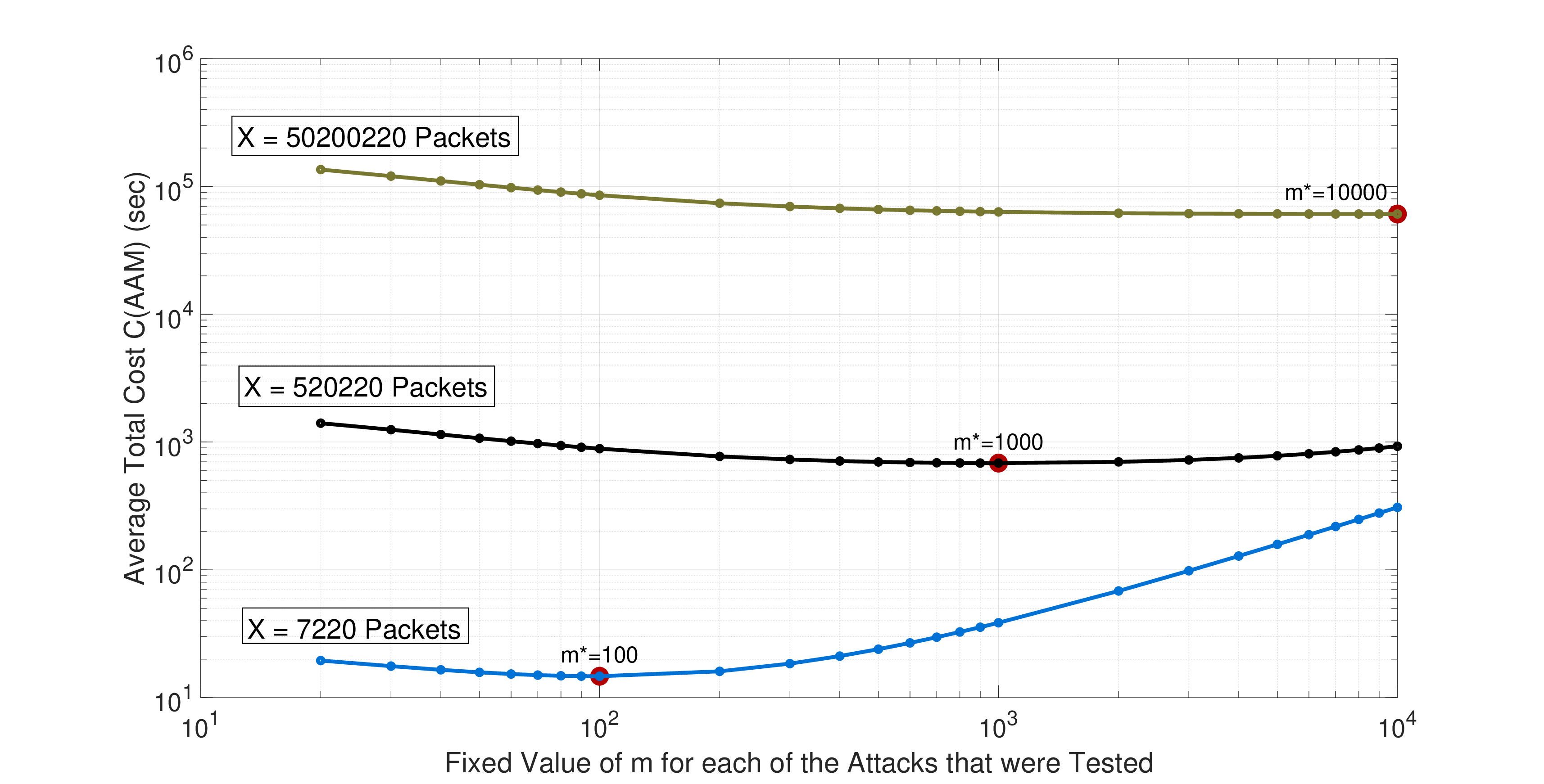}
	\caption{Theoretical graph of the Average Total Cost plotted against the parameter $m$ when both the $x$ and $y$ axes are logarithmic for $\frac{\beta}{\alpha}=0.05$.  We show how the Average Cost $C(AAM)$ depends on $m$ and exhibit the theoretical minimum Average Total Cost at the value $m^*$, which is obtained analytically.}
	\label{fig:optimizationAAM}
\end{figure}

In this subsection, we present the measurements of the proposed AAM scheme. Figure \ref{fig:optimizationAAM} illustrates the theoretically computed average total cost $C(AAM)$, across various parameter values when  $W=20$, $\alpha = 20\beta$, and we vary the value of $E[X]$. We have run experiments where we first generate the value of  $X$, i.e. the number of packets in an attack against the Gateway Server, at random. These experiments are run thirty times for each fixed value of the average number of packets $E[X]$ received during a Flood Attack. The experiments are also run for several different values of $E[X]$. In each experiment, we compute the corresponding value of the cost function $C(AAM)$ that results from the attack and compute its average value by taking the average of the cost values obtained from the distinct measurements for a given $E[X]$. 

\begin{figure}[h!]
	\hspace*{-21pt}
	\centering
	\includegraphics[height=6cm,width=9.5cm]{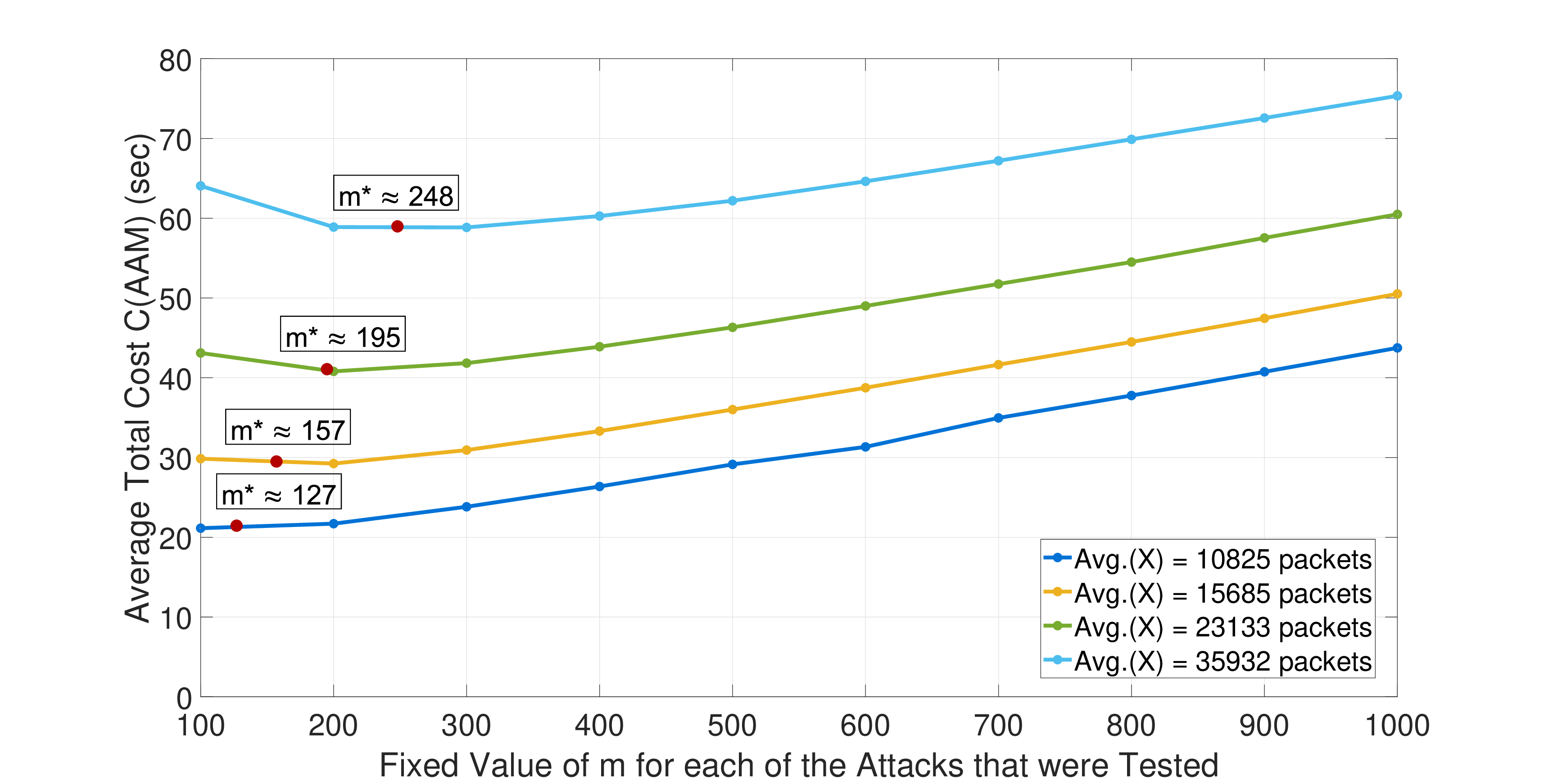}
	\caption{Experimental graph (when both the $x$ and $y$ coordinates are linear, i.e. not logarithmic) of the Average Total Cost $C(AAM)$ over thirty independent experiments for $\frac{\beta}{\alpha}=0.05$, plotted against the parameter $m$. We see that the average cost depends on $m$, and confirms that the theoretically predicted $m^*$ provides a useful value of cost minimization.}
	\label{AverageOverhead}
\end{figure}

These experimental results are summarized in
Figure \ref{AverageOverhead}. They provide an empirical demonstration of the accuracy and validity of the theoretical formula (\ref{opt-m}) which predicts the value of $m^*$.

\subsection{The Effect of the AAM on the Gateway's Time-Line Behaviour during Attacks}

We now measure  the time dependent queue length of packets at the entrance of the SQF during two successive Flood Attacks. The first attack involves over $10,000$ packets, and the second one is comprised of  some $40,000$ packets. The AD system processes the incoming packet stream and provides an ATTACK alert, which is sent to the AAM. The AAM then measures the SQF input queue length, and  uses the formula (\ref{opt-m}), to choose $m^*$, which is $127$ for the first attack, and $248$ for the second one. 

\begin{figure}[h!]
	\hspace*{-27pt}
	\centering
	\includegraphics[height=6cm,width=10.2cm]{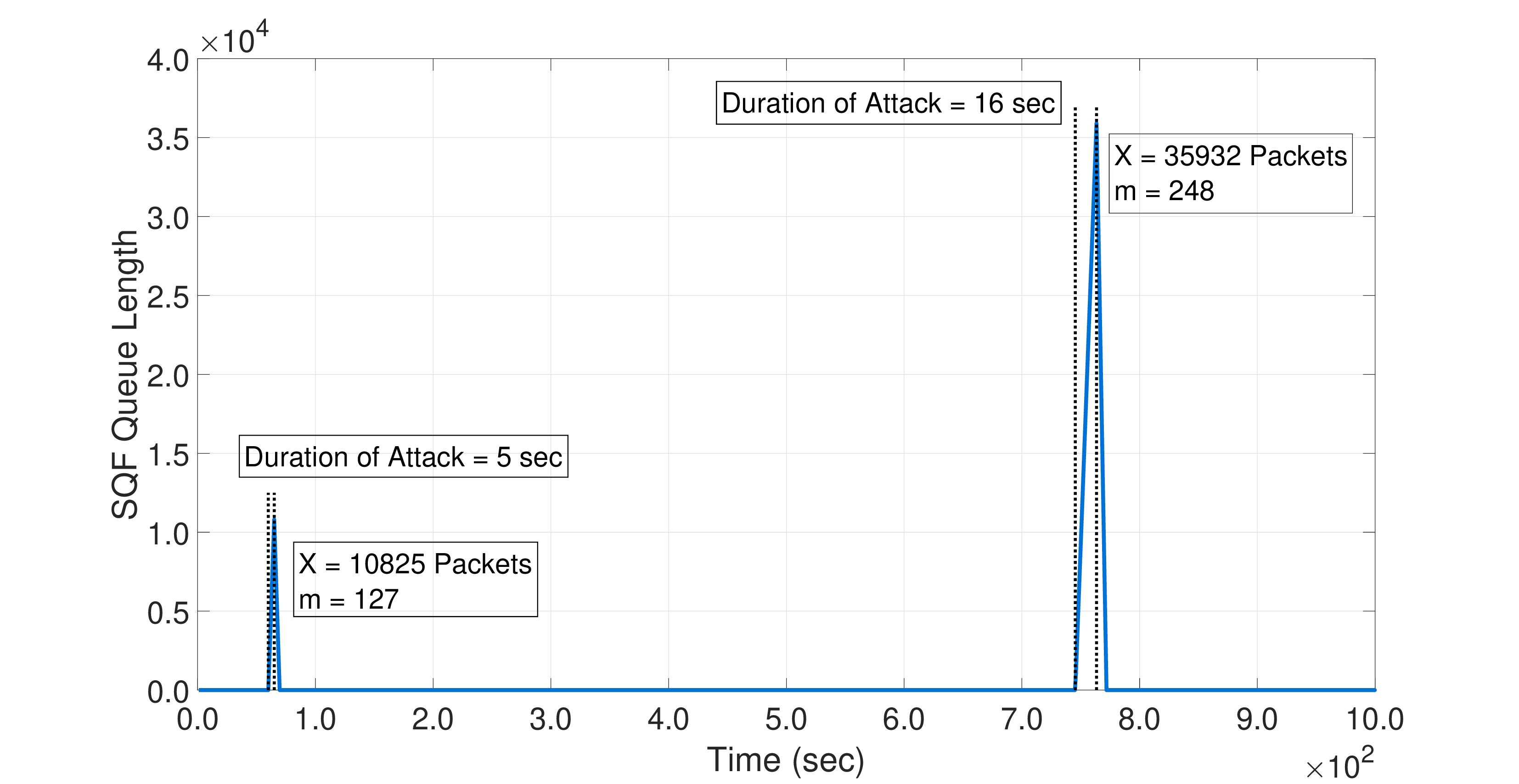}
	\caption{Time-line of queue length measurements at the entrance of the SQF for two successive attacks. The first  attack involves around $10,000$ packets, while the second one is more severe and involves close to $40,000$ packets. The AD processes the incoming traffic from the SNMP in batches of $W$ packets, and provides an attack alert when an attack is detected. AAM then goes into action, observes the packet lengths, and uses the formula (\ref{opt-m}) to select the optimum
		value $m^*$, which is $127$ in the first case, and $248$ in the second case. We observe the effectiveness of the AAM in rapidly dropping attack packets, and  also reacting to the second attack in a timely fashion after the first attack is cleared.}
	\label{ExtraOverheadSQF}
\end{figure}

The effectiveness of the AAM is demonstrated  in Figure \ref{ExtraOverheadSQF}, by the speed with which the AAM drops packets after each of the two successive attacks, and reacts to the second attack in a timely fashion despite the first attack.

\begin{figure}[h!]
	\hspace*{-27pt}
	\centering
	\includegraphics[height=6cm,width=10.2cm]{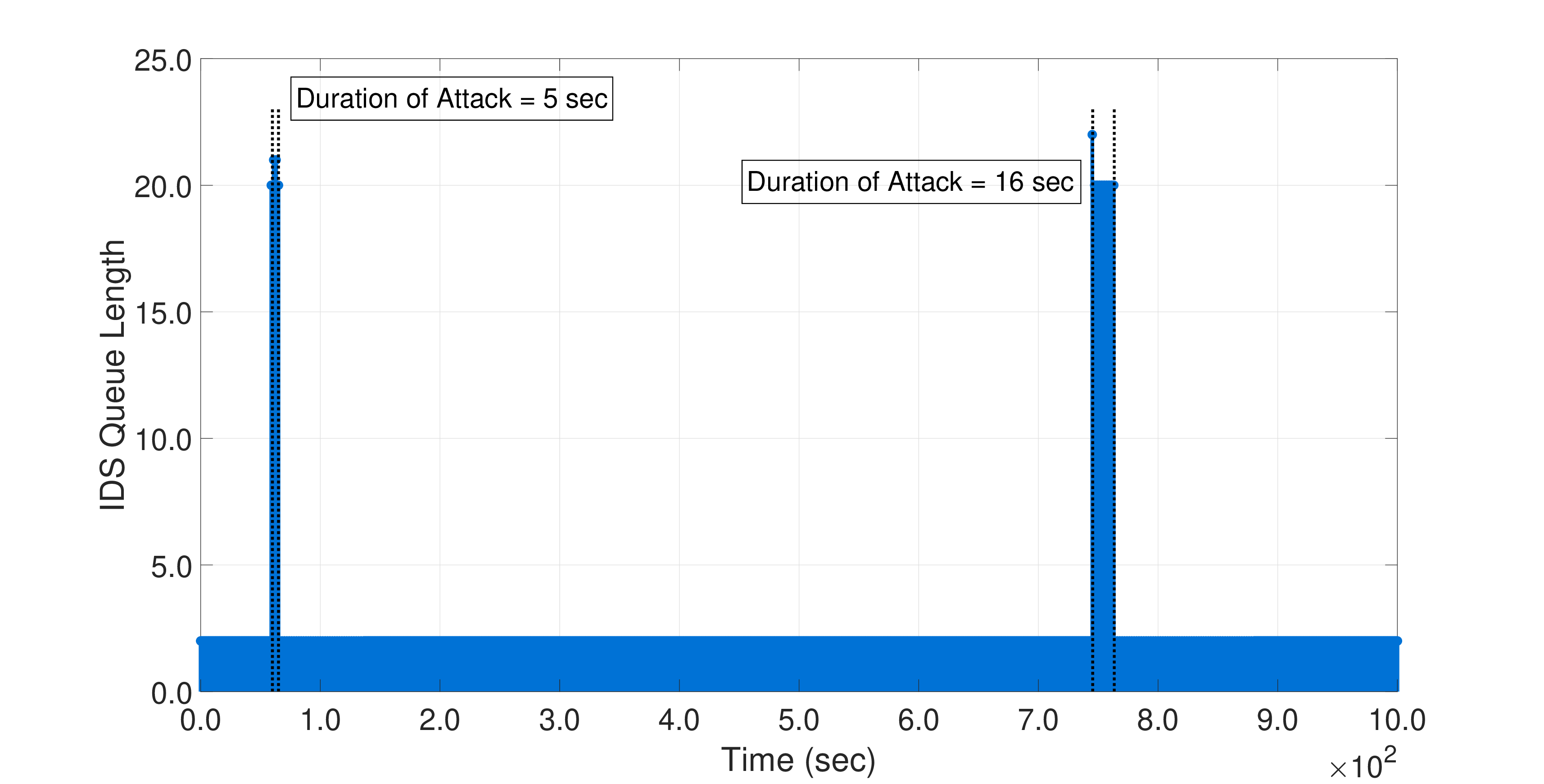}
	\caption{Time-line of queue length measurements at the entrance of the AD system, when the two successive attacks shown in Figure \ref{ExtraOverheadSQF} occur. Despite the fact that
		the two successive attacks involve $10,000$ and $40,000$ packets, the conjunction of the use of the SQF and the AAM, limits the AD input queue to around $20$ packets for a very short period, without overwhelming the AD and allowing it to operate smoothly and continuously.}
	\label{NoOverheadIDS}
\end{figure}

For the two attacks that were shown in Figure \ref{ExtraOverheadSQF}, the time dependent packet queue length at the entrance of the AD system is shown in Figure \ref{NoOverheadIDS}. This latter figure also demonstrates that the joint use of the SQF and the AMM, is able to 
effectively limit the AD input queue to a very small number of $20$, very rapidly after the attack begins;
thus, the 
AD system is able  to operate continuously and effectively, contributing to the correct decisions that are being made by the AAM.

\section{Conclusions} \label{Conc}

In this research, we have first studied the impact of UDP Flood Attacks on an IoV Gateway Server that supports an AD  system.  We demonstrate that even short-duration attacks can cause significant overload for the Sever. the Gateway Server. As a result, The Server's normal operations, including AD, are substantially slowed down. Indeed, we observe that an attack that lasts $60$ seconds may create a backlog of packets at the Server that requires several hours to clear out.

Thus, we propose that the Gateway Server's input should be ``protected''  by a special SQF front end that operates the QDTP policy, to allow the timely operation of the Server even when an attack occurs. This approach requires 
installing an inexpensive lightweight  hardware addition, such as an RPi, 
between the local area network that supports the Sensors,  and the IoV Gateway Server. 

Several experiments are then used to illustrate the effectiveness of the proposed approach, and we note that the SQF front end requires that a key timing parameter $D$ be chosen. We provide a theoretical analysis of how $D$ should be selected, and show that it must be comparable but larger than the AD processing time per packet  under non-attack conditions. We validate this theoretical observation with several experiments and show that the SQF effectively preserves the Gateway Server from congestion and overload, allowing it to operate normally, even in the presence of severe Flood Attacks. 

However, we note that the congestion that is eliminated at the Server may now accumulate at the SQF input, creating excessive packet processing work for the Gateway.  
Therefore, we propose the novel mitigation action AAM which is activated when the AD detects an attack. The AAM  is able to drop incoming attack packets in a relatively short time, and is designed to minimize a cost function that combines the number of dropped benign packets and the overhead of
testing  the incoming packet stream for AD to identify the end of the attack. The AAM is tested experimentally and shown to be very effective during significant Flood Attacks. 

While this paper has focused on an architecture with multiple sources of IoV traffic and a single Gateway, future work will consider Edge Systems with multiple devices and Gateways. We plan to study dynamic policies for AD for complex  IoV networks having both static and mobile nodes, with mitigation techniques that include traffic routing and packet drops at the Edge.  

The energy consumption of such systems is another important issue that we plan to address, so that dynamic management policies may minimize energy consumption, optimize system Quality of Service, and offer Cybersecurity.

\section*{Acknowledgement}
This work was partially funded by the EU Horizon Program Grant Agreement No. 101120270 for the DOSS research project.
		
\bibliographystyle{IEEEtran}
\bibliography{IDS,datasets,RNN,references_federated,references,references2,references3,references4,mybib,security_issues_references,references_arnn_conference,IoVSecurity_references}

	\end{document}